\documentclass[aps,pre,twocolumn,showpacs,superscriptaddress]{revtex4-1}

\usepackage[utf8]{inputenc}
\usepackage[T1]{fontenc}
\usepackage[german,english]{babel}
\usepackage{amsmath}
\usepackage{mathtools}
\usepackage{amssymb}
\usepackage{bbm}
\usepackage{graphicx}
\usepackage{hyperref}
\usepackage[arrow, matrix, curve]{xy}
\usepackage{bm}
\usepackage{yhmath}
\usepackage{color}
\usepackage{url}
\usepackage{mathrsfs}
\usepackage{verbatim}
\usepackage{tabularx}
\usepackage{textcomp}
\usepackage{booktabs}
\usepackage{siunitx}

\usepackage[usenames,dvipsnames]{xcolor}
\hypersetup{colorlinks=true, linkcolor=BrickRed, urlcolor=blue!50!black, citecolor=blue!50!black}

\usepackage[normalem]{ulem}
\makeatletter
\newcommand\hide@visible[1]{%
  \bgroup\fboxsep=.3ex\colorbox{Gray}{begin hide}%
  #1\colorbox{Gray}{end hide}\egroup%
}
\newcommand\hide@hidden[1]{%
  \bgroup\fboxsep=.3ex\colorbox{Gray}{hidden text}%
}
\newcommand\hide@invisible[1]{}
\newcommand\makevisible{\let\hide\hide@visible}
\newcommand\makehidden{\let\hide\hide@hidden}
\newcommand\makeinvisible{\let\hide\hide@invisible}
\makeatother
\makehidden

\begin{document}

%\preprint{}

\title{Thermophoretic motion of a charged single colloidal particle}



%\author{Autoren}
%\affiliation{Institut f\"ur Theoretische Physik, Universit\"at Erlangen-N\"urnberg, Staudtstra{\ss}e~7, 91058, Erlangen, Germany}

\author{Daniel B. Mayer} 
\affiliation{Institut f{\"u}r Theoretische Physik,  Universit\"at Innsbruck,
Technikerstra{\ss}e 21A, A-6020 Innsbruck, Austria}

\author{Dieter Braun}
\affiliation{Systems Biophysics, Physics Department, Nanosystems Initiative Munich and Center for NanoScience, Ludwig-Maximilians-Universit{\"a}t M{\"u}nchen, Amalienstra{\ss}e 54, 
D-80799 M{\"u}nchen, Germany}

\author{Thomas Franosch} 
\affiliation{Institut f{\"u}r Theoretische Physik,  Universit\"at Innsbruck,
Technikerstra{\ss}e 21A, A-6020 Innsbruck, Austria}

% \email[]{thomas.franosch@uibk.ac.at}
%\thanks{}
\date{\today}

\begin{abstract}
We calculate the thermophoretic drift of a charged single colloidal particle with hydrodynamically slipping surface immersed in an electrolyte solution in response to a small temperature gradient. Here, we rely on a linearized hydrodynamic approach for the fluid flow and the motion of the electrolyte ions while keeping the full nonlinearity of the Poisson-Boltzmann equation of the unperturbed system to account for possible large surface charging. The partial differential equations are transformed into a coupled set of ordinary differential equations in linear response. Numerical solutions are elaborated for parameter regimes of small and large Debye shielding and different hydrodynamic boundary conditions encoded in a varying slip length. Our results are in good agreement with predictions from recent theoretical work and successfully describe experimental observations on thermophoresis of DNA. We also compare our numerical results with experimental data on polystyrene beads.
\end{abstract}

\maketitle

\section{Introduction}

Nonequilibrium transport processes of charged colloids or macromolecules in aqueous solutions are ubiquitous in biological, chemical and physical systems~\cite{Georgis:EPC:1998, Mast:PRL_104:2010, Rainard:SLAS_23:2018, Prieve:PNAS_116:2019, Battat:Softmatter_15:2019, Reichl:JACS_136:2014}. Typically, the motion of such colloids is mediated by externally maintained thermodynamic (bulk) gradients mostly in solute concentration, electric potential and temperature. The phoretic motion then depends in a subtle manner on the surface properties of the colloid and its interactions with the solvent whose details are still subject of ongoing scientific research, experimentally~\cite{Piazza:PRL_88:2002, Duhr:PNAS_103:2006, Ning:Langmuir_24:2008, Reichl:PRL_112:2014, Syshchyk:EPJE_39:2016, Battat:Softmatter_15:2019, Prieve:PNAS_116:2019, Burelbach:JCP_147:2017} as well as theoretically~\cite{
Anderson:ARFM_21:1989, Parola:EPJE_25:2004, Dhont:Langmuir_23:2007, Khair:PF_21:2009, Wuerger:PRL_101:2008, Hill:PRSA_471:2015, Burelbach:EPJE_42:2019, Shin:POF_32:2020, Kocherginsky:JCP_154:2021}.

In particular, the directed drift motion in response to a temperature gradient, usually referred to as thermophoresis is a formidable problem due to its peculiar sensitivity on the details of the system under investigation. It depends not only on particle properties such as molecular weight~\cite{Wuerger:PRL_102:2009}, size~\cite{Putnam:Langmuir_23:2007, Braibanti:PRL_100:2008, Duhr:PRL_96:2006, Duhr:PNAS_103:2006}, anisotropy~\cite{Gardin:PCCP_21:2019, Gittus:EPJE_42:2019}, concentration~\cite{Burelbach:EPJE_41:2018}, surface charging, and surface coating~\cite{Ning:Langmuir_24:2008}, but also on solvent parameters including permittivity, salinity, Debye screening length, and thermoelectric field, as well as their inherent temperature dependence~\cite{Wuerger:PRL_101:2008, Morthomas:EPJE_27:2008, Duhr:PRL_96:2006, Reichl:PRL_112:2014}. For example, already the dependence of the thermophoretic drift velocity on the dimensions of the colloid has been observed differently for the same system under investigation. While the study in Ref.~\cite{Duhr:PNAS_103:2006} suggests a linear variation with particle size, measurement data from Refs.~\cite{Putnam:Langmuir_23:2007, Braibanti:PRL_100:2008} strongly supports a constant thermophoretic drift motion of the particle.

This results in competing contributions to thermophoretic transport rendering it more complex to understand and predict than other field-driven transport processes such as electrophoresis or diffusiophoresis. Nevertheless, thermophoresis has numerous (bio-) technological and microfluidic applications, for example, it plays a pivotal role for the separation and characterization of polymers and macromolecules by thermal field-flow fractionation~\cite{Giddings:Science_260:1993}, the trapping and enrichment of DNA in a microchannel with ambient flow~\cite{Duhr:PRL_96:2006, Duhr:PRL_97:2006}, the possible guiding of fluid motion by thermal micropumps~\cite{Tan:PRA_11:2019, Semenov:EPJE_42:2019}, and in the state-of-the-art analysis of biomolecular interactions by means of microscale thermophoresis (MST)~\cite{JarabekWillemsen:JMS_1077:2014}.

Thermally driven transport was first observed by the Irish physicist John Tyndall in aerosols by simple noticing that a temperature gradient affects the motion of dust particles tending to avoid hot surfaces~\cite{Tyndall:SA:1870}. Shortly afterwards, the German physiologist Carl Ludwig discovered a similar effect in aqueous alkali halide solutions in 1856~\cite{Ludwig:KAW_20:1856}, which then was independenly considered in detail by the  Swiss physico-chemist Charles Soret in 1879~\cite{Soret:JPTA_9:1879}. The phenomenon is therefore also called the Ludwig-Soret effect or just Soret effect. 

In principle, thermophoresis of a charged colloidal particle immersed in an aqueous electrolyte solution constitutes a highly nonlinear transport problem coupling ion convection-diffusion dynamics, electrostatics, and solvent flow. This makes a quantitative analysis of the underlying field equations and their corresponding boundary conditions within a continuum approach almost intractable. Nevertheless, most studies regard thermophoresis more or less explicitly as a linear-response phenomenon~\cite{Wuerger:PRL_101:2008, Morthomas:EPJE_27:2008, Rasuli:PRL_101:2008}, where the equilibrium electrolyte structure around the colloid is only slightly distorted by the applied temperature gradient.
Then to linear order, the thermophoretic drift velocity of the colloidal particle becomes
\begin{equation}
\mathbf{U}_{T}=-D_{T}\nabla T,
\end{equation}
with $D_{T}$ being referred to as the thermal diffusion coefficient which may take both signs indicating that the colloid migrates to the cold for positive $D_{T}$ and to the warm for negative values, respectively. This transport coefficient constitutes an Onsager cross-coefficient relating heat and particle flux within the framework of non-equilibrium thermodynamics~\cite{DeGroot:NET:1984, Burelbach:EPJE_41:2018}.
Considering symmetry arguments, the linearized set of partial differential equations can be significantly simplified and thus the problem of calculating the thermal diffusion coefficient $D_{T}$ essentially reduces to finding a solution to a coupled set of ordinary differential equations with suitable boundary conditions, similar to the treatment of the problem of electrophoresis by O'Brien and White~\cite{OBrien:JCS_74:1978}.
While in the two limiting cases of thin and wide Debye layers, the scale disparity as well as weak surface charging allow for approximate analytic solutions~\cite{Morthomas:EPJE_27:2008, Wuerger:PRL_101:2008}, a numerical approach is generally neccessary to capture the subtle interplay of the underlying  transport mechanisms for the full range of parameters. Here the focus lies on the response of the aqueous electrolyte to the temperature gradient, in particular, how concentration gradients in the bulk solution and the accompanying thermoelectric potential affect the thermal transport coefficient via boundary conditions. Furthermore, effects arising from a strong surface charging can be properly revealed only by retaining the full nonlinear Poisson-Boltzmann equation governing the equilibrium potential instead of applying its linearized form in Debye-H{\"u}ckel approximation valid only for weakly charged particles.
 
Based on these considerations, we provide here a comprehensive review of the thermophoresis problem of a charged spherical colloid within linear response following the theoretical approach of Rasuli and Golestanian~\cite{Rasuli:PRL_101:2008}. Moreover, for completeness, we also discuss in detail the correct representation of the electrolyte bulk behavior in terms of suitable far-field boundary conditions since this has been paid little attention to in literature so far, but seems to be crucial  to correctly determine the thermal diffusion coefficient. Comparison with other most recent theoretical work on thermophoresis~\cite{Burelbach:EPJE_42:2019} supports our explanations. 
This paper is organized as follows: In Sec.~\ref{sec:Theory}, we reformulate the generic thermophoresis problem within a hydrodynamic continuum approach. Then the linear response of the system is addressed in Sec.~\ref{Sec:linear_response_theory}, where we derive the relevant linear differential equations for thermophoretic transport, while Sec.~\ref{Sec:diff_contributions_thermophoresis} provides a short discussion of different contributions to the thermal diffusion coefficient. In Secs.~\ref{Sec:Decomposition_problem}-\ref{Sec:nondimensonal_problem}, we elaborate the techniques to considerably simplify these differential equations relying on strategies originally introduced by O'Brien and White~\cite{OBrien:JCS_74:1978} to tackle the electrophoresis problem. In the following Sec.~\ref{Sec:Numerical_procedure}, the solution procedure to obtain numerical solutions to the ordinary differential equations is described, while in Secs.~\ref{Sec:Rasuli_model}-\ref{Sec:PSB} the results are compared to other theoretical approaches as well as to experimental data on thermophoresis of DNA and polystyrene beads. Last, we conclude in Sec~\ref{Sec:Summary_conclusion}.

\section{Theory}\label{sec:Theory}

In this section, we introduce a minimal theoretical continuum model for a charged single colloid in an aqueous electrolyte solution exposed to a stationary and spatially uniform temperature gradient. Here a description in terms of field equations is employed, where the behavior of the bulk solution is accounted for by suitable far-field boundary conditions.
In particular, we elaborate the linear response of the system to small temperature gradients in order to calculate the thermal diffusion coefficient for arbitrary Debye layer width and possibly large surface charging. 

Most theoretical approaches to thermophoresis of colloids discussed in the literature~\cite{Anderson:ARFM_21:1989, Wuerger:RPP_73:2010, Wuerger:PRL_101:2008, Morthomas:EPJE_27:2008, Rasuli:PRL_101:2008} constitute extensions of the theory of electrophoresis~\cite{OBrien:JCS_74:1978, Mangelsdorf:JCS_88:1992, Ohshima:ACIS_62:1995, Khair:PF_21:2009, Hill:PRSA_471:2015, Schmitz:IOP_24:2012, Schnitzer:PRE_86:2012}. 
Our theoretical description follows the same path. In particular, when concerning the solution strategy of the corresponding field equations using asymptotic expressions for the relevant quantities, we strongly rely on the techniques of O'Brien and White in their seminal work on electrophoresis~\cite{OBrien:JCS_74:1978}.

\subsection{Formulation of the thermophoresis problem}

The system of interest is a  charged chemical inert dielectric spherical particle, immersed in a large  electrolyte reservoir, where the completely ionized solute consists of $N$ different ionic species of charge $z_{i}e$ with elementary charge $e$ and valences $z_{i}$ $(i=1,2,\ldots,N)$.   This reservoir can exchange heat with the surroundings and at the boundary a thin charged layer emerges due to ionic density gradients setting up a thermoelectric field (see Appendix~\ref{App:Soret_effect_bulk}). 
At the interface between solid and electrolyte, a Debye double layer of characteristic width $1/\kappa$ forms, screening the surface charge of the colloidal particle. It comprises a thin immobile layer of adsorbed counter-charged ions on the solid surface adjacent to an otherwise diffusive cloud of mobile ions~\cite{Lyklema:FICS_2:1995}.
The double layer connects smoothly to an electroneutral bulk region  within the electrolyte-domain boundary.
Then a stationary and spatially uniform temperature gradient $\nabla T$ is  applied externally, resulting in a phoretic motion of the neutrally buoyant spherical particle with steady-state velocity $\mathbf{U}_{T}$ relative to the quiescent electrolyte. This drift motion  is a consequence of the local hydrodynamic stresses in the surrounding solution~\cite{Anderson:ARFM_21:1989} induced by gradients in ion concentrations and electric potential (see Appendix \ref{App:Soret_effect_bulk}) in the bulk solution, as well as the corresponding temperature-induced asymmetry of the Debye double layer. 

In addition to the Debye length, a second length scale is characteristic for the system, namely the distance from the particle center to the  hydrodynamic slipping plane~\cite{Delgado:JCIS_309:2007}. The solvent inside may remain attached to the particle surface and a hydrodynamically stagnant layer builds up, except for a small region of slip length $\lambda$~\cite{Lauga:SHEFM:2007} accounting for the possible hydrophilic or hydrophobic nature of the particle surface~\cite{Gopmandal:SM_17:2021, Park:Electrophoresis_34:2013}. Thus, the slipping plane can be understood as the effective or virtual boundary of the colloidal particle with hydrodynamic radius $a$, where the electrolyte is assumed to be unaffected by the applied temperature gradient.
In the remainder, we employ a reference frame attached to the center of the colloidal particle. Hence, in the far field the solvent flow approaches a uniform stream $-\mathbf{U}_{T}$ and within the slipping plane the velocity is zero.
The accompanied temperature profile is assumed to change only linearly in the temperature gradient
\begin{equation}\label{Eq:temperature_field}
T(\mathbf{r})=T_{0}+\mathbf{r} \cdot \nabla T,
\end{equation}
where $T_{0}$ denotes the reference temperature in the center of the spherical particle. The presence of the colloidal particle does not alter the applied temperature gradient since thermal conductivities of the solvent and the core material of the colloid are assumed to be comparable. In contrast, for metallic particles the local temperature variations around the colloid may be of central importance~\cite{Giddings:JCIS_176:1995}. 
Furthermore, the ions are treated as non-interacting  particles, dispersed in a fluid that consists mainly of solvent molecules, yielding an ideal dilute solution. These assumptions justify a continuum description of the thermophoresis problem, where the colloid is considered as a macroscopic object compared to the solutes and the surrounding solvent as a dielectric continuous medium~\cite{Brady:JFM_667:2011}.
The fundamental equations governing thermophoretic transport in terms of the electrostatic potential $\phi(\mathbf{r})$, the ion concentration $n_{i}(\mathbf{r})$ for each species $i=1,2,\ldots, N$, the pressure $p(\mathbf{r})$ and the velocity field $\mathbf{u}(\mathbf{r})$ within a stationary state, are presented in the following.

\subsubsection{Governing field equations}

The Poisson equation relates the electrostatic potential outside the colloidal particle to the free charge density
\begin{equation}
 \rho(\mathbf{r})=\sum_{i=1}^{N}z_{i} e n_{i}(\mathbf{r}),
\end{equation}
via
\begin{equation}\label{Eq:Poisson_full}
 \nabla\cdot\left[\epsilon_{\text{r}}(\mathbf{r})\nabla\phi(\mathbf{r})\right]=-\frac{1}{\epsilon_{0}}\rho(\mathbf{r}),
\end{equation}
where the space dependence of the relative dielectric permittivity $\epsilon_{\text{r}}(\mathbf{r})$ is inherited from the thermal gradient, since the permittivity depends on temperature. 
 Here $n_{i}(\mathbf{r})$ denotes the local concentration of the ions and $\epsilon_{0}$ is the vacuum permittivity. 

The  current density of the ionic solutes is phenomenologically modified along the lines of Onsager's linear response relation between conjugate fluxes and forces~\cite{DeGroot:NET:1984} and reads
\begin{align}\label{Eq:current_full}
 \mathbf{j}_{i}(\mathbf{r})&= n_{i}(\mathbf{r})\mathbf{u}(\mathbf{r})-\mu_{i}^{0}z_{i} e n_{i}(\mathbf{r})\nabla\phi(\mathbf{r}) \nonumber \\
&\quad -D_{i}(\mathbf{r})n_{i}(\mathbf{r})\left[S_{T}^{i}\nabla T(\mathbf{r})+\nabla\log n_{i}(\mathbf{r})\right].
\end{align}
It accounts for the combined effects of advection, electric migration, as well as thermal and mass diffusion, where $D_{i}(\mathbf{r})=\mu_{i}^{0}k_{\text{B}}T(\mathbf{r})$ denotes the Einstein diffusion coefficients evaluated at the local temperature. Thus the assumption is that the ion mobilities $\mu_{i}^{0}$ are temperature-independent and the Stokes-Einstein relation holds locally. The \emph{ionic Soret coefficients} $S_{T}^{i}$ of the salt cations and anions comprises the thermophoretic response of the solutes due to hydration by surrounding water molecules~\cite{Helfand:JCP_32:1960, Agar:JPC_93:1989, Takeyama:JSC_17:1988} and a thermoelectric field~\cite{Guthrie:JCP_17:1949, Burelbach:EPJE_41:2018, Wuerger:RPP_73:2010}  acting on the ions (see Appendix \ref{App:Soret_effect_bulk}). In principle, these Soret coefficients could also be temperature-dependent, however we shall be interested only in the effects linear in the temperature gradient. Consequently we can evaluate them at the reference temperature $T_{0}$. In the following, they are treated as known input parameters.

In the stationary state, the currents are source-free and satisfy the extended Nernst-Planck equations
\begin{equation}\label{Eq:Nernst_Planck_full}
\nabla\cdot\mathbf{j}_{i}(\mathbf{r})=0.
\end{equation}
In addition, we consider the momentum-balance equation for the solvent and shall neglect  effects of inertia in the limit of small Reynolds number. It is known as the stationary Stokes equation for a Newtonian fluid
\begin{equation}\label{Eq:Stokes_full}
\nabla P(\mathbf{r})-\eta\nabla^{2}\mathbf{u}(\mathbf{r})=\mathbf{f}^{\text{el}}(\mathbf{r}),
\end{equation}
 accompanied by the incompressibility constraint 
\begin{equation}\label{Eq:Incombressible_full}
\nabla\cdot\mathbf{u}(\mathbf{r})=0.
\end{equation}
The electric body force density is obtained as
\begin{equation}\label{Eq:Helmholtz_force_density}
\mathbf{f}^{\text{el}}(\mathbf{r})=-\rho(\mathbf{r})\nabla\phi(\mathbf{r})-\frac{\epsilon_{0}}{2}\mathbf{E}(\mathbf{r})^2\nabla\epsilon_{\text{r}}(\mathbf{r})
\end{equation}
from the divergence of the Korteweg-Helmholtz stress tensor for an electrically linear dielectric material~\cite{Helmholtz:AdP_249:1881, Korteweg:AdP_245:1880, Landau:ECM:2013}.
The first term on the right-hand side (r.h.s.) of Eq.~\eqref{Eq:Helmholtz_force_density} denotes the electrostatic force density while the second is a dielectric contribution accounting for the polarization of the solvent in the local electric field $\mathbf{E}(\mathbf{r})=-\nabla\phi(\mathbf{r})$. For an incompressible solvent, the electrostrictive contribution due to variations in the relative dielectric permittivity with respect to the solvent mass density $\rho_{\text{m}}$ as well as the hydrostatic pressure can be absorbed in an effective pressure~\cite{Saville:ARFM_29:1997} 
\begin{equation}
P(\mathbf{r})=p(\mathbf{r})-\frac{\epsilon_{0}}{2}\rho_{\text{m}}\mathbf{E}(\mathbf{r})^{2}\left(\frac{\partial\epsilon_{\text{r}}(\mathbf{r})}{\partial\rho_{\text{m}}}\right)_{T} .
\end{equation}
Here $p(\mathbf{r})$ denotes the hydrodynamic pressure and $\eta$ is the  viscosity of the solvent.
We ignore effects arising from a possible temperature dependence of the viscosity.

\subsubsection{Boundary conditions}

At the stationary (virtual) surface of the colloidal particle with hydrodynamic radius $a$, the boundary conditions are specified by means of the unit normal $\mathbf{n}$ pointing into the solvent. Then, by virtue of the electric Gauss law, the electric displacements in both the dielectric particle and the solvent are connected to the effective surface charge density $\sigma(\mathbf{r})$ by
\begin{equation}\label{Eq:boundary_phi_full}
\left[ \epsilon_{\text{r}}(\mathbf{r}) \frac{\partial}{\partial n}\phi(\mathbf{r})-\epsilon_{\text{r}}^{\text{in}}(\mathbf{r}) \frac{\partial}{\partial n}
\phi^{\text{in}}(\mathbf{r})\right]\bigg|_{r=a}=-\frac{\sigma(\mathbf{r})}{\epsilon_{0}},
\end{equation}
where $\epsilon_{\text{r}}^{\text{in}}(\mathbf{r})$ is the dielectric permittivity of the core material and $\partial/\partial n=\mathbf{n}\cdot\nabla$ denotes the normal derivative at the surface. In principle, the potential inside the particle $\phi^{\text{in}}(\mathbf{r})$ has to be obtained from Laplace's equation $\nabla\cdot\left[\epsilon^{\text{in}}_{\text{r}}(\mathbf{r})\nabla\phi^{\text{in}}(\mathbf{r})\right]=0$, together with the continuity condition $(\phi(\mathbf{r})-\phi^{\text{in}}(\mathbf{r}))|_{r=a}=0$. However, the ratio of the dielectric permittivities is small for the particles of interest~\cite{Saville:ARFM_9:1977}, such that we can neglect contributions from the electric field inside the particle. 

Furthermore, the electrolyte solution within the region between the solid particle surface and the slipping plane is assumed  to be unaffected neither by the applied temperature gradient nor by the accompanied electric field and displays no macroscopic motion. Consequently, electrochemical reactions, mostly from dissociation of surface functional groups or adsorption of ions and surface conduction~\cite{Mangelsdorf:JCSFT_86:1990, Carrique:JCIS_227:2000, Carrique:JCIS_243:2001} due to possible lateral motion within the slipping plane, are absent, yielding a radially symmetric surface-charge density $\sigma_{0}$ on the colloidal particle independent of the temperature.
Then Eq.~\eqref{Eq:boundary_phi_full} simplifies to
\begin{equation}\label{Eq:boundary_phi_simpl}
\frac{\partial\phi(\mathbf{r})}{\partial n}\bigg|_{r=a}=-\frac{\sigma_{0}}{\epsilon_{0}\epsilon_{\text{r}}(\mathbf{r})}.
\end{equation}
Under these conditions, the ion currents together with the velocity normal to the particle vanish
\begin{subequations}
\begin{align}
\mathbf{n} \cdot \mathbf{j}_{i}(\mathbf{r})|_{r=a}&=0, \label{Eq:Impenetrable_bound_cond_flux}\\
\mathbf{n} \cdot \mathbf{u}(\mathbf{r})|_{r=a}&=0, \label{Eq:Impenetrable_bound_cond_velo}
\end{align}
\end{subequations}
since ions cannot penetrate the slipping plane. 
The velocity obeys a Navier boundary condition~\cite{Navier:ARS_6:1823}
\begin{equation}\label{Eq:Navier_bound_cond}
\mathbf{u}_{\text{t}}(\mathbf{r})|_{r=a}=\frac{\lambda}{\eta}\left[\boldsymbol{\sigma}^{\prime}(\mathbf{r}) \cdot \mathbf{n}-\left(\mathbf{n}\cdot\boldsymbol{\sigma}^{\prime}(\mathbf{r})\cdot \mathbf{n}\right)\mathbf{n}\right]|_{r=a},
\end{equation}
linearly relating the tangential component of the electrolyte velocity $\mathbf{u}_{\text{t}}(\mathbf{r})=\mathbf{u}(\mathbf{r})-\left(\mathbf{u}(\mathbf{r})\cdot\mathbf{n}\right)\mathbf{n}$ to the shear stress tensor $\boldsymbol{\sigma}^{\prime}(\mathbf{r})=\eta\left[\nabla\mathbf{u}(\mathbf{r})+\left(\nabla\mathbf{u}(\mathbf{r})\right)^{\text{T}}\right]$ at the slipping plane~\cite{Lauga:SHEFM:2007}. Here $\lambda$ denotes the  slip length, which we treat as a known input parameter. For $\lambda=0$ the usual no-slip boundary condition is recovered. 

At large distances away from the colloidal particle within the electroneutral bulk region (not yet in the vicinity of the electrolyte domain boundary), the electric field approaches the  thermoelectric field
as a consequence of the thermoelectric force $\mathbf{F}_{i}=z_{i} e \mathbf{E}^{\text{th}}$ directly acting on the  ions~\cite{Guthrie:JCP_17:1949, Burelbach:EPJE_42:2019}. 
 To linear order in the thermal gradient the thermoelectric field is uniform
\begin{equation}\label{Eq:bound_phi_full}
 \lim\limits_{|\mathbf{r}|\rightarrow\infty}{\nabla\phi(\mathbf{r})}=-\mathbf{E}^{\text{th}} = \phi^{\text{th}} \frac{\nabla T}{T_0},
\end{equation}
where the response coefficient $\phi^{\text{th}}$ is referred to as the thermoelectric potential (see Appendix~\ref{App:Soret_effect_bulk}). 

 Furthermore, the ion concentrations approach their bulk behavior arising from the redistribution of the salt ions~\cite{Wuerger:RPP_73:2010} due to the temperature gradient. To linear order in the thermal gradient
 (Appendix~\ref{App:Soret_effect_bulk}), the ion concentrations behave asymptotically for $|\mathbf{r}| \to \infty$ as 
\begin{align}\label{Eq:bound_conc_full}
 n_{i}(\mathbf{r})&\sim n_{i}^{\text{b}}(\mathbf{r}) = n_{i,0}^{\text{b}} \left[1- S_T^i\, \mathbf{r} \cdot \nabla T  \right]. 
\end{align}
 This is a striking difference to other phoretic transport processes, such as diffusiophoresis~\cite{Keh:Langmuir_16:2000} or electrophoresis~\cite{Burelbach:EPJE_42:2019}, since there one avoids the interdependence of companion fields in the bulk, whereas in thermophoresis, the inherent coupling of the thermoelectric field and the gradient in ion concentrations has to be accounted for (see especially Eq.~\eqref{Eq:density_gradient_bulk} in Appendix \ref{App:Soret_effect_bulk}).

Finally, we have to specify the far-field stream velocity
\begin{equation}\label{Eq:far_field_velocity}
\lim\limits_{|\mathbf{r}|\rightarrow\infty}{\mathbf{u}(\mathbf{r})}=-\mathbf{U}_{T} ,
\end{equation}
by the requirement for phoretic motion, that the total force acting on the colloidal particle vanishes~\cite{Brady:JFM_667:2011}. There is no need to include a zero-torque constraint, as the problem displays axial symmetry. Here $\mathbf{U}_{T}$ denotes the thermophoretic velocity attained by the particle under steady-state conditions. The calculation of its magnitude $|\mathbf{U}_{T}|$ constitutes the goal of our investigations.

\subsection{Linear-response theory}\label{Sec:linear_response_theory}

We are solely interested in the linear response of the system to an externally applied temperature gradient. Correspondingly   relative temperature changes over distances of the order of the extend of the colloid including its Debye layer, $a+\kappa_{0}^{-1}$, are considered to be small as characterized by the following condition
\begin{equation} 
\left(a+\kappa_{0}^{-1}\right)\frac{|\nabla T|}{T_{0}}\ll 1.
\end{equation}
Here the inverse (equilibrium) Debye screening length $\kappa_{0}$ is defined via
\begin{equation}
\kappa_{0}^{2}=\frac{1}{ k_B T_0\epsilon_{0}\epsilon_{\text{r}}^{0}}\sum_{i=1}^{N}z_{i}^{2}e^2 n_{i,0}^{b},
\end{equation}
with dielectric permittivity $\epsilon_{\text{r}}^{0}$ and constant bulk ion concentration $n_{i,0}^{\text{b}}$ evaluated at the reference temperature $T_{0}$.
In this case, the electrical double layer is only slightly distorted from its equilibrium configuration by the applied temperature gradient and the subsequent particle motion. 
This allows linearizing the governing nonlinear partial differential equations, together with the corresponding boundary conditions, in the perturbation with respect to the  spherically symmetric reference state, which corresponds  to thermal equilibrium with a uniform temperature $T_{0}$, such that no solvent flow $\mathbf{u}_{0}=0$ occurs. Consequently, we can write the field variables within linear response as
\begin{subequations}\label{Eq:perturb_expansion}
\begin{align}
\mathbf{u}(\mathbf{r})&=\delta\mathbf{u}(\mathbf{r}), \\
%T(\mathbf{r})&=T_{0}+\delta T(\mathbf{r}), \\
 n_{i}(\mathbf{r})&=n_{i}^{0}(r)+\delta n_{i}(\mathbf{r}),\\
 P(\mathbf{r})&=P_{0}(r)+\delta P(\mathbf{r}),  \\
 \phi(\mathbf{r})&=\phi_{0}(r)+\delta \phi(\mathbf{r}),
\end{align}
\end{subequations}
where $n_{i}^{0}(r), P_{0}(r)$ and $\phi_{0}(r)$ denote the reference quantities with $r=\lvert\mathbf{r}\rvert$ and the perturbation terms are proportional to $\lvert\nabla T\rvert$ to lowest order.
%, since $\delta T(\mathbf{r})=\mathbf{r} \cdot \nabla T$ (see Eq.~\eqref{Eq:temperature_field}). 
The thermophoretic velocity is thus linearly related to the weak temperature gradient by
\begin{equation}\label{Eq:Def_therm_diff}
\mathbf{U}_{T}=-D_{T}\nabla T,
\end{equation}
defining the thermal diffusion coefficient as $D_{T}$. Consequently the calculation of $\lvert\mathbf{U}_{T}\rvert$ to linear order in the temperature gradients is equivalent  to  determining $D_{T}$. 

\subsubsection{Reference system}

Substituting now the expansion [Eqs.~\eqref{Eq:perturb_expansion}] into the nonlinear field equations [Eqs.~\eqref{Eq:Poisson_full} and \eqref{Eq:Nernst_Planck_full}-\eqref{Eq:Incombressible_full}], we arrive to zeroth order in the perturbation at the equilibrium electrokinetic equations
\begin{subequations}
\begin{align}
 0&=\nabla^{2}\phi_{0}(r)+\frac{1}{\epsilon_{0}\epsilon_{\text{r}}^{0}}\rho_{0}(r), \label{Eq:Poisson_equ} \\
 0&=\nabla P_{0}(r)+\rho_{0}(r)\nabla \phi_{0}(r), \label{Eq:Stokes_zero} \\
 0&=\nabla\cdot\left[\mu_{i}^{0}z_{i} e n_{i}^{0}(r)\nabla\phi_{0}(r) +D_{i}^{0}\nabla n_{i}^{0}(r)\right], \label{Eq:continuity_equ_zero}
\end{align}
\end{subequations}
with charge density $\rho_{0}=\sum_{i}z_{i}e n_{i}^{0}(r)$ and spatially uniform diffusion coefficients $D_{i}^{0}=\mu_{i}^{0} k_{\text{B}} T_0$.
A solution for the continuity equation [Eq.~\eqref{Eq:continuity_equ_zero}] exists for vanishing  fluxes, $\mathbf{j}_{i}^{0}(r)=0$, recovering the Boltzmann distribution
\begin{equation}\label{Eq:Boltzmann_distr}
n_{i}^{0}(r)=n_{i,0}^{\text{b}}\exp\left[- \frac{ z_{i}e \phi_{0}(r)}{k_{\text{B}}T_0}\right],
\end{equation}
where the potential vanishes in the electroneutral bulk,
\begin{equation}\label{Eq:bound_phi_zero_limes}
 \lim\limits_{r\rightarrow\infty}{\phi_{0}(r)}=0.
\end{equation}
Inserting now this ion distribution into Eq.~\eqref{Eq:Poisson_equ} and using the spherical symmetry yields the nonlinear Poisson-Boltzmann equation
\begin{equation}\label{Eq:Poisson_Boltzmann_full}
\frac{1}{r^2}\frac{\textrm{d}}{\textrm{d}r}\left[r^2\frac{\textrm{d}}{\textrm{d}r}\phi_{0}(r)\right]=-\frac{\rho_{0}(r)}{\epsilon_{0}\epsilon_{\text{r}}^{0}},
\end{equation}
determining the overall electrostatic potential~\cite{Debye:Phys_Z_24:1923}. The corresponding boundary condition [Eq.~\eqref{Eq:boundary_phi_simpl}] reduces to 
\begin{equation}
\frac{\textrm{d}\phi_{0}(r)}{\textrm{d}r}\bigg|_{r=a}=-\frac{\sigma_{0}}{\epsilon_{0}\epsilon_{\text{r}}^{0}}.
\end{equation}
Furthermore, a local balance between pressure gradients and electric body forces [Eq.~\eqref{Eq:Stokes_zero}] maintains a spherically symmetric solvent distribution around the colloidal particle with local solute (osmotic) pressure
\begin{equation}
 P_{0}(r)=k_{\text{B}}T_0 \sum_{i=1}^{N}\left(n_{i}^{0}(r)-n_{i,0}^{b}\right),
\end{equation}
and vanishing pressure  at infinity, $P_{0}(r)\to 0$ as $r\rightarrow\infty$.

\subsubsection{Linearized equations}

Retaining only first-order perturbation terms, a set of coupled linear field equations is obtained:
\begin{subequations}\label{Eq:Gen_Electrokinetic_Eq_lin}
\begin{align}
 0&=\nabla^{2}\delta\phi(\mathbf{r})+\frac{1}{\epsilon_{0}\epsilon_{\text{r}}^{0}}\delta\rho(\mathbf{r}) \nonumber \\
 &\quad-\alpha\nabla\cdot\left(\frac{\mathbf{r} \cdot \nabla T
}{T_{0}}\nabla\phi_{0}(r)\right), \label{Eq:Poisson_lin} \\
0&=\eta \nabla^{2}\mathbf{u}(\mathbf{r})-\nabla\delta P(\mathbf{r}) -\rho_{0}(r)\nabla\delta\phi(\mathbf{r}) \nonumber \label{Eq:Stokes_lin}  \\
   &\quad-\delta\rho(\mathbf{r})\nabla\phi_{0}(r)+\frac{1}{2}\alpha\epsilon_{0}\epsilon_{\text{r}}^{0}[\nabla\phi_{0}(r)]^{2}\frac{\nabla T}{T_{0}}, \\
 0&=\nabla\cdot\left[n_{i}^{0}(r)\mathbf{u}(\mathbf{r}) - D_i^0 n_i^0(r) \nabla \left(\frac{\delta n(\mathbf{r})}{n_i^0(r)}\right) \right. \nonumber\\
 &\left.\qquad\quad-\mu_{i}^{0} z_{i} e n_{i}^{0}(r)\nabla\delta\phi(\mathbf{r})-D_{i}^{0}n_{i}^{0}(r)S_{T}^{i}\nabla T\right. \nonumber \\
  &\left.\qquad\quad  -D_{i}^{0}\frac{ 
\mathbf{r} \cdot \nabla T}{T_{0}} \nabla n_{i}^{0}(r)\right], \label{Eq:Nernst_Planck_lin}
  \end{align}
\end{subequations}
with charge density variations $\delta\rho(\mathbf{r})=\sum_{i}z_{i} e\delta n_{i}(\mathbf{r})$. Here, gradients in the dielectric permittivity have been evaluated as $\nabla\epsilon_{\text{r}}(\mathbf{r})=-\alpha\epsilon_{\text{r}}^{0}\nabla T/T_{0}$ by expanding the dielectric permittivity in temperature gradients
\begin{equation}
 \epsilon_{\text{r}}(\mathbf{r})=\epsilon_{\text{r}}^{0}+\left(\frac{\partial\epsilon_{\text{r}}}{\partial T}\right) \mathbf{r} \cdot \nabla T =\epsilon_{\text{r}}^{0}-\frac{\alpha\epsilon_{\text{r}}^{0}}{T_{0}}\mathbf{r} \cdot \nabla T ,
\end{equation}
with logarithmic derivative $\alpha=-\partial\ln\epsilon_{\text{r}}/\partial\ln T$.

This set of generalized electrokinetic equations [Eqs.~\eqref{Eq:Gen_Electrokinetic_Eq_lin}] for thermophoresis, requires the solution of the  full nonlinear Poisson-Boltzmann equation [Eq.~\eqref{Eq:Poisson_Boltzmann_full}] as input. 
In principle, these coupled partial differential equations constitute a possible starting point for theoretical investigations of thermophoresis. 
However, to streamline the further analysis, we follow Ref.~\cite{Rasuli:PRL_101:2008} and introduce a set of ionic potential functions 
\begin{align}\label{Eq:ionic_potential}
 \Omega_{i}(\mathbf{r})=&\frac{\delta n_{i}(\mathbf{r})}{n_{i}^{0}(r)}+S_{T}^{i} \mathbf{r} \cdot \nabla T+\frac{z_{i} e\delta\phi(\mathbf{r})}{k_{\text{B}}T_0} \nonumber \\
 &-\frac{z_{i} e\left(\phi_{0}(r)+\phi^{\text{th}}\right)}{k_{\text{B}}T_0} \frac{\mathbf{r} \cdot \nabla T}{T_{0}}, 
\end{align}
which is suggested from the linearization of a Boltzmann-type ansatz
\begin{align}
 n_{i}(\mathbf{r})=n_{i,0}^{b}&\exp\left[-\frac{z_{i} e\phi(\mathbf{r}) }{k_{\text{B}} T(\mathbf{r})}+\Omega_{i}(\mathbf{r})\right. \nonumber \\
&\left.\quad\quad-S_{T}^{i}\mathbf{r}\cdot\nabla T(\mathbf{r})- \frac{z_{i} e \mathbf{r} \cdot \mathbf{E}^{\text{th}}}{k_{\text{B}} T(\mathbf{r})}\right],
\end{align}
for the ion concentrations. Here the first term in the exponential is of a local-equilibrium form, the last two terms anticipate the thermophoretic motion of the ionic solutes in the bulk (see Appendix \ref{App:Soret_effect_bulk}) and $\Omega_{i}(\mathbf{r})$ parametrizes the residual genuine nonequilibrium effects. 
Then Eq.~\eqref{Eq:Nernst_Planck_lin} yields
\begin{align}\label{Eq:Omega_spelled_out}
 &\nabla^{2}\Omega_{i}(\mathbf{r})- \frac{z_{i} e \nabla\phi_{0}(r)}{k_{\text{B}} T_0}\cdot \left[ \nabla\Omega_{i}(\mathbf{r}) - \frac{\mathbf{u}(\mathbf{r})}{D_{i}^{0}}\right]= \nonumber \\
 &\quad- \frac{z_{i} e \nabla\phi_{0}(r)}{k_{\text{B}} T_0} \cdot \left(1- \frac{z_{i} e [\phi_{0}(r)+ \phi^{\text{th}}]}{k_{\text{B}} T_0} \right)\frac{\nabla T}{T_{0}},
\end{align}
after spelling out the divergence. Gradients in the perturbed pressure $\delta P(\mathbf{r})$ and the electrostatic potential $\delta\phi(\mathbf{r})$ are eliminated by taking the curl of Eq.~\eqref{Eq:Stokes_lin}, leading to
\begin{align}\label{Eq:Stokes_lin_decoupled}
 &\eta\nabla^2(\nabla \times\mathbf{u}(\mathbf{r}))  -\sum_{i=1}^{N}z_{i} e n_{i}^{0}(r)\nabla\Omega_{i}(\mathbf{r})\times\nabla\phi_{0}(r) = \nonumber \\
 &\sum_{i=1}^{N}z_{i} e n_{i}^{0}(r)\left[ \frac{z_{i} e ( \phi_{0}(r)+ \phi^{\text{th}})}{k_{\text{B}} T_0} -S_{T}^{i}T_{0}\right]\frac{\nabla T}{T_{0}}\times\nabla\phi_{0}(r) \nonumber \\
&-\frac{1}{2}\alpha\epsilon_{0}\epsilon_{\text{r}}^{0} \nabla  |\nabla \phi_0(r)|^2  \times\frac{\nabla T}{T_{0}}.
\end{align}
The introduction of the potential function $\Omega_{i}(\mathbf{r})$ considerably simplifies the task of computing the thermal diffusion coefficient $D_{T}$, since it decouples Eqs.~\eqref{Eq:Stokes_lin} and~\eqref{Eq:Nernst_Planck_lin} from the Poisson Eq.~\eqref{Eq:Poisson_lin}. Note that the r.h.s.\@ of Eqs.~\eqref{Eq:Omega_spelled_out} and\eqref{Eq:Stokes_lin_decoupled} depend  (nonlinearly) on the reference system, while the dependence on the unknowns $\Omega_i(\mathbf{r})$ and $\mathbf{u}(\mathbf{r})$ on the left-hand side (l.h.s.) is linear by construction. 

To obtain a complete specification of the thermophoresis problem, it still remains to determine the boundary conditions for the perturbed field quantities $\mathbf{u}(\mathbf{r})$ and $\Omega_{i}(\mathbf{r})$. At the colloidal surface, we impose the Navier condition for the solvent velocity [Eqs.~\eqref{Eq:Navier_bound_cond} and \eqref{Eq:Impenetrable_bound_cond_velo}], together with a vanishing radial ion current [Eqs.~\eqref{Eq:Impenetrable_bound_cond_flux}], yielding within linear response
\begin{equation}
\frac{\partial \Omega_{i}(\mathbf{r})}{\partial n}\bigg|_{r=a}+ \frac{z_{i} e [\phi_{0}(r)+\phi^{\text{th}}]}{k_{\text{B}} T_0} \frac{\mathbf{n} \cdot \nabla T}{T_{0}}\bigg|_{r=a}  =0.
\end{equation}
In the far field, the velocity obeys Eq.~\eqref{Eq:far_field_velocity} to lowest order in the temperature gradients.
Furthermore, by means of Eqs.~\eqref{Eq:bound_phi_full} and~\eqref{Eq:bound_phi_zero_limes} the perturbed potential behaves  asymptotically as
\begin{equation}\label{Eq:Boundary_pert_pot}
 \delta\phi(\mathbf{r})\sim \phi^{\text{th}} \frac{ \mathbf{r}\cdot \nabla T}{T_0}  \quad\text{for}\quad |\mathbf{r}|\rightarrow\infty,
\end{equation}
as a consequence of the thermoelectric migration [Eq.~\eqref{Eq:Electric_field_bulk}]. In addition, according to Eq.~\eqref{Eq:bound_conc_full} the perturbation in ion concentrations should tend to
\begin{equation}\label{Eq:Boundary_pert_den}
 \frac{\delta n_{i}(\mathbf{r})}{n_{i,0}^{b}}\sim -S_{T}^{i} \,\mathbf{r} \cdot \nabla T \quad\text{for}\quad |\mathbf{r}|\rightarrow\infty,
\end{equation}
arising from the gradients in bulk concentration. 
Hence, it follows from Eq.~\eqref{Eq:ionic_potential} that we have to impose
\begin{equation}\label{Eq:Omega_bound_infty}
 \lim\limits_{|\mathbf{r}|\rightarrow\infty}{\Omega_{i}(\mathbf{r})}=0 ,
\end{equation}
within the bulk region. These boundary conditions together with the corresponding Eqs.~\eqref{Eq:Omega_spelled_out} and~\eqref{Eq:Stokes_lin_decoupled} enable us to calculate the response of the isolated colloidal particle to the small temperature gradient and its accompanying fields. In the next sections we shall show that the asymptotic behavior of the functions $\mathbf{u}(\mathbf{r)}$ and $\Omega_{i}(\mathbf{r)}$ completely determines the linear response, i.e.\@ the thermal diffusion coefficient.
The linearized Poisson equation [Eq.~\eqref{Eq:Poisson_lin}] is redundant.

\subsection{Different Contributions to thermophoresis}\label{Sec:diff_contributions_thermophoresis}

In principle, the thermal diffusion coefficient $D_{T}$ is determined by four contributions. The first is due to the electrostatic energy density of the different ionic solutes within the temperature-induced asymmetric Debye double layer~\cite{Burelbach:EPJE_42:2019} and is represented by the term $\propto \sum_{i}^{N}z_{i}^{2} e^2 n_{i}^{0}(r) \phi_{0}(r)$ in Eq.~\eqref{Eq:Stokes_lin_decoupled}. A second stems from polarization effects of the solvent in the local electric field and can be interpreted as hydration enthalpy density~\cite{Landau:ECM:2013, Burelbach:EPJE_41:2018, Burelbach:EPJE_42:2019}. It corresponds to the last term on the r.h.s. \@ of Eq.~\eqref{Eq:Stokes_lin_decoupled}. The last two contributions originate from the thermophoretic behavior of the ions in the bulk solution encoded in the term $\propto \sum_{i}^{N}z_{i} e n_{i}^{0}(r) [z_i e  \phi^{\text{th}}/k_{\text{B}}T_0- S_{T}^{i}T_{0}] $ and the far-field boundary conditions 
[Eqs.~\eqref{Eq:Boundary_pert_pot} and~\eqref{Eq:Boundary_pert_den}]. We refer to it as ion hydration effect. More specifically, we define the contribution arising from the boundary condition for the disturbed electrostatic potential only as electrophoretic contribution to ion hydration, as it is reminiscent of the electrophoresis problem.

Since the field equations for the perturbed fields are linear, we can disentangle the different contributions by discarding inhomogeneities or changing the far-field boundary conditions.  
For example, the electrostatic contribution is obtained by keeping in Eq.~\eqref{Eq:Stokes_lin_decoupled}
only the relevant energy-density terms and imposing the far-field boundary conditions
\begin{subequations}
\begin{align}
\lim\limits_{|\mathbf{r}|\rightarrow\infty}{\delta\phi(\mathbf{r})}=0, \\
\lim\limits_{|\mathbf{r}|\rightarrow\infty}{\frac{\delta n_{i}(\mathbf{r})}{n_{i,0}^{\text{b}}}}=0.
\end{align}
\end{subequations}
Similarly, by retaining the original boundary conditions and artificially switching off the relevant terms related to the electrostatic energy, contributions from ion and colloid hydration can be compared. 

\subsection{Decomposition of the problem}\label{Sec:Decomposition_problem}

The appearance of the thermal diffusion coefficient $D_{T}$ in the far-field boundary condition [Eq.~\eqref{Eq:far_field_velocity}] for the velocity makes the problem of solving the governing generalized electrokinetic equations intricate.
Using the technique of O' Brien and White~\cite{OBrien:JCS_74:1978}, we circumvent this difficulty by exploiting the linearity of the derived field equations together with the corresponding boundary conditions and writing the overall solution as a superposition of the solutions for the following two simpler auxiliary problems: 
\begin{itemize}
 \item[(1)] The spherical particle held fixed in a flow field $-\mathbf{U}$ in the absence of any applied temperature gradient $\nabla T$ yielding the far-field boundary conditions
 			\begin{subequations}
            \begin{align}
             \lim\limits_{|\mathbf{r}|\rightarrow\infty}{\mathbf{u}(\mathbf{r})}&=-\mathbf{U}, \label{Eq:Bound_cond_U_1} \\
             \lim\limits_{|\mathbf{r}|\rightarrow\infty}{\delta\phi(\mathbf{r})}&=0, \\
             \lim\limits_{|\mathbf{r}|\rightarrow\infty}{\frac{\delta n_{i}(\mathbf{r})}{n_{i,0}^{b}}}&=0.
            \end{align}
            \end{subequations}
\item[(2)] The spherical particle held fixed in a temperature gradient $\nabla T$ in a quiescent electrolyte far away from the colloidal particle with far-field boundary conditions
			\begin{subequations}
            \begin{align}
             \lim\limits_{|\mathbf{r}|\rightarrow\infty}{\mathbf{u}(\mathbf{r})}&=0, \label{Eq:Bound_cond_U_2} \\
             \delta\phi(\mathbf{r})&\sim -\mathbf{r} \cdot \mathbf{E}^{\text{th}} \quad\text{for}\quad |\mathbf{r}|\rightarrow\infty, \\
             \frac{\delta n_{i}(\mathbf{r})}{n_{i,0}^{b}}&\sim -S_{T}^{i}\mathbf{r} \cdot \nabla T\quad\text{for}\quad |\mathbf{r}|\rightarrow\infty.
            \end{align}
            \end{subequations}
\end{itemize}
The sum of the solutions to the Eqs.~\eqref{Eq:Omega_spelled_out} and~\eqref{Eq:Stokes_lin_decoupled} for each of these problems then satisfies the desired far-field boundary condition [Eq.~\eqref{Eq:Omega_bound_infty}].
Concomitantly, we have to ensure the constraint that for thermophoretic motion the net force acting on the particle is zero~\cite{Brady:JFM_667:2011}.
Within linear response, the forces required to hold the colloidal particle fixed for each problem read 
\begin{subequations}
\begin{align}
\mathbf{F}^{(1)}&=\gamma^{(1)}\mathbf{U}, \\
\mathbf{F}^{(2)}&=\gamma^{(2)} \frac{\nabla T}{T_0},
\end{align}
\end{subequations}
where $\gamma^{(1)}$ and $\gamma^{(2)}$ are constants to be determined. 
The superposition of the forces gives then rise to a vanishing  net force $\mathbf{F}=\mathbf{F}^{(1)}+\mathbf{F}^{(2)}=0$, provided we choose
\begin{equation}
 \mathbf{U}=-\frac{\gamma^{(2)}}{\gamma^{(1)}} \frac{\nabla T}{T_0}.
\end{equation}
Thus, by comparison with Eq.~\eqref{Eq:Def_therm_diff}  the thermal diffusion coefficient is read off as
\begin{equation}\label{Eq:D_T_force}
 D_{T}=\frac{\gamma^{(2)}}{\gamma^{(1)}} \frac{1}{T_0}.
\end{equation}
Furthermore, this method yields also the  diffusion coefficient $D=k_{\text{B}} T_0/\gamma^{(1)}$ of a charged spherical particle from the solution to problem (1).

\subsection{Symmetry considerations}\label{Sec:Symmetry}

\begin{figure}[htbp]
\centering
\includegraphics[width=0.48\textwidth]{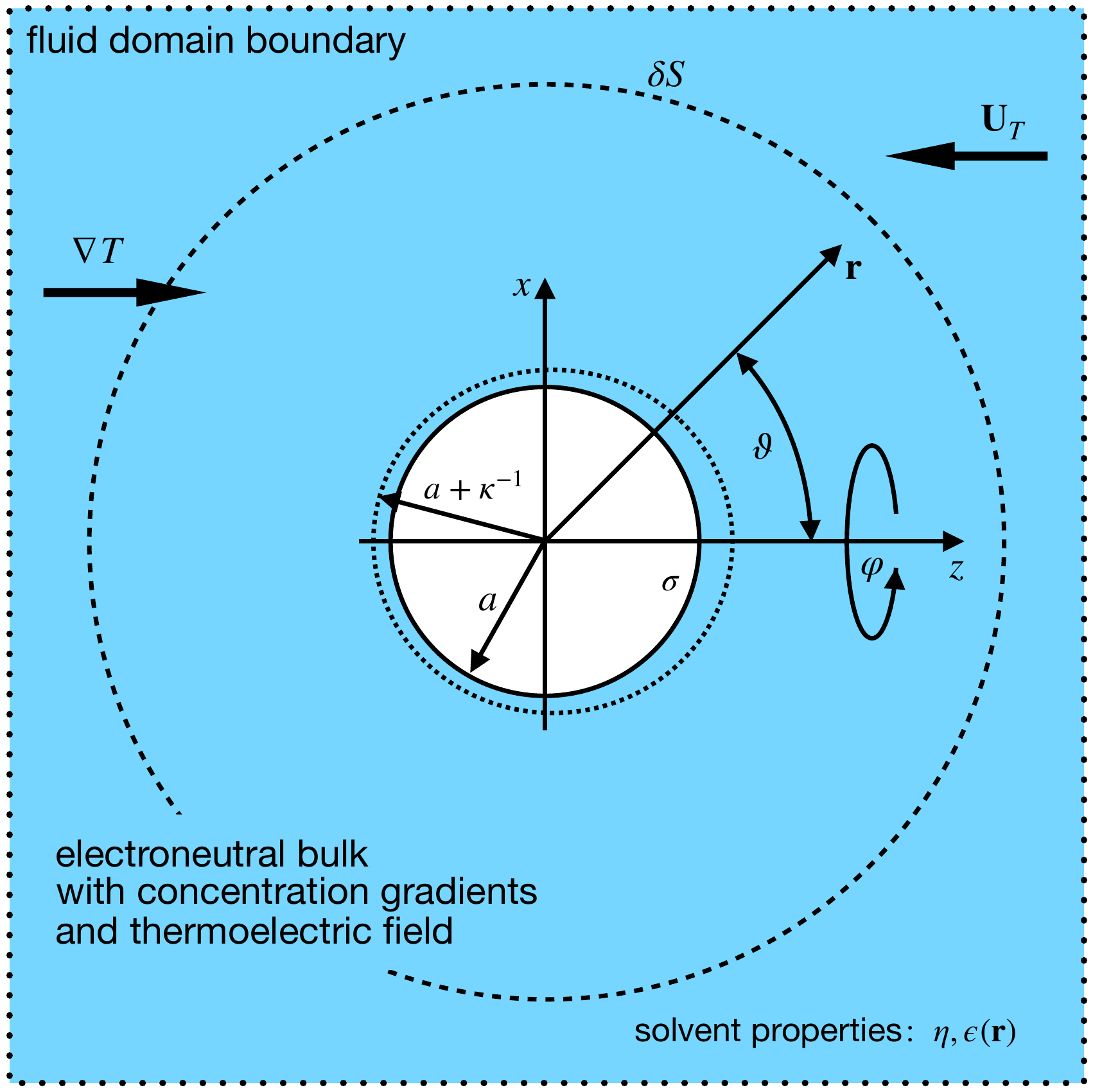}
\caption{Schematics  (not to scale) of the colloidal particle with hydrodynamic radius $a$ carrying a surface charge density $\sigma$ in a particle-fixed reference frame. A small temperature gradient $\nabla T$ is applied from outside. The short-dashed line denotes the outer edge of the  slightly distorted Debye double layer of width $1/\kappa$, while the dotted line corresponds to the fluid-domain boundary located at a macroscopic distance from the colloid. The integration boundary $\delta S$ in the electroneutral bulk is shown as big-dotted line. The solvent displays a spatially varying dielectric constant $\epsilon(\mathbf{r})$ due to the temperature gradient.}
\label{fig:plot_1}
\end{figure}

The reference system without gradients exhibits spherical symmetry, while both auxiliary problems display only axial symmetry due to the imposed perturbations either by the flow $\mathbf{U}$ or the thermal gradient $\nabla T$. We choose the origin of the coordinate system to be at the center of the colloid and the $z$-direction to be aligned with the flow, respectively with the thermal gradient (see Fig.~\ref{fig:plot_1}).  
Thus the temperature is  represented as
\begin{equation}
 T(\mathbf{r})=T_0 + \mathbf{r} \cdot  \nabla T =T_0 + \lvert\nabla T\rvert r\cos\vartheta.
\end{equation}
Furthermore both auxiliary problems (1) and (2) are discussed in parallel by introducing 
\begin{equation}
 \mathbf{X}=\begin{dcases}
             \mathbf{U}, & (1) \\
             \nabla T /T_0, & (2).
            \end{dcases}
\end{equation}
To linear order in $\mathbf{X}$ all scalar potentials are then of the form 
$ f(r) (\mathbf{r}/r)\cdot \mathbf{X} $ with some spherically symmetric function $f(r)$, while no pseudo-scalar fields can be constructed.  
Accordingly, a convenient representation of the solenoidal velocity field is introduced by
\begin{equation}\label{Eq:Debye_represenation}
\mathbf{u}(\mathbf{r})=\nabla\times\left[\mathbf{r}\psi(\mathbf{r})\right]-\nabla\times\left[\mathbf{r}\times\nabla\chi(\mathbf{r})\right] ,
\end{equation}
in terms of two scalar functions, called toroidal $\psi(\mathbf{r})$ and poloidal Debye potential $\chi(\mathbf{r})$~\cite{Gray:AJP_46:1978}. Owing to the fact that no pseudo-scalar fields arise within linear response with respect to $\mathbf{X}$, the velocity fields can be written as 
\begin{subequations}\label{Eq:Velocity_field_sym}
\begin{align}
\mathbf{u}(\mathbf{r})&=-\nabla\times\left[\mathbf{r}\times\nabla\chi(\mathbf{r})\right]-\mathbf{U}, \quad &(1) \\
\mathbf{u}(\mathbf{r})&=-\nabla\times\left[\mathbf{r}\times\nabla\chi(\mathbf{r})\right]. \quad &(2) 
\end{align}
\end{subequations}
Finally, we express the ion potentials and the poloidal Debye potential as 
\begin{subequations}\label{Eq:Debye_symmetry}
\begin{align}
\Omega_{i}(\mathbf{r}) &=\omega_{i}(r)(\mathbf{r} /r)\cdot\mathbf{X}, \\
\chi(\mathbf{r})& =R(r)(\mathbf{r} /r)\cdot\mathbf{X} ,
\end{align}
\end{subequations}
with radially symmetric unknowns $\omega_i(r)$ and $\chi(r)$ 
for each of the two problems. 

With these symmetry-adapted forms for $\mathbf{u}(\mathbf{r})$ and $\Omega_{i}(\mathbf{r})$, the linearized partial differential equations [Eqs.~\eqref{Eq:Omega_spelled_out} and~\eqref{Eq:Stokes_lin_decoupled}] reduce to a set of coupled linear ordinary differential equations, drastically simplifying the task of calculating the thermal diffusion coefficient.

\subsection{Calculating the force acting on the colloid}\label{Sec:Force_calc}

In order to obtain the thermal diffusion coefficient, we first have to determine the forces acting on the colloidal particle for each problem (1) and (2). A common procedure is to integrate viscous and electrical traction forces over the surface of the spherical particle relying on the calculation of gradients in the potential and velocity. However, we avoid this cumbersome procedure following again a method suggested by O'Brien and White~\cite{OBrien:JCS_74:1978} for the electrophoresis problem and compute the forces from the asymptotic form of the velocity field $\mathbf{u}_{\text{as}}(\mathbf{r})$ behind the Debye double layer in the bulk solution. This is possible, since in the momentum balance equation neither inertial terms no body forces enter, rather all forces derive from a stress tensor. Thus, by Gauss' theorem the total force on the colloid is the same as the  total force on any concentric sphere containing the colloid.  
At large radii, this force will be only due to the viscous drag, since forces due to electric fields either rapidly decay or cancel upon integrating over the sphere. 
Another convenient aspect of this approach is that it does not require computing the disturbances in the potential $\delta\phi(\mathbf{r})$. 

Hence, we consider a large sphere $S$ enclosing the particle and the Debye double layer. Its radius has been taken sufficiently large in order to enclose the region where the charge density $\rho(\mathbf{r})$  is non-negligible, since in the bulk solution  local charge neutrality holds (see Fig.~\ref{fig:plot_1}). Consequently the \emph{total} electric force on the combined system becomes zero and the external forces $\mathbf{F}$ for problems (1) and (2) are counterbalanced by integrating viscous traction forces over the surface $\partial S$ of the sphere,
\begin{equation}
 \mathbf{F}=-\int_{\partial S}\boldsymbol{\sigma}(\mathbf{r})\cdot\mathbf{n}\,\textrm{d}S,
\end{equation}
where
\begin{equation}
 \boldsymbol{\sigma}(\mathbf{r})=-P(\mathbf{r})   \mathbb{I}+\eta\left[\nabla\mathbf{u}(\mathbf{r})+(\nabla\mathbf{u}(\mathbf{r}))^{T}\right] ,
\end{equation}
denotes the viscous stress tensor for the respective problems. 

Next, we show how the velocity and pressure fields behave asymptotically for large distances and calculate the corresponding forces. The electric forces decay rapidly  in Eq.~\eqref{Eq:Stokes_lin} as $r\rightarrow\infty$ leading to a simplified momentum balance equation 
\begin{equation}\label{Eq:Force_free_Stokes}
-\eta \nabla\times\nabla\times\mathbf{u}(\mathbf{r})-\nabla\delta P(\mathbf{r})=0,
\end{equation}
together with the corresponding boundary conditions [Eqs.~\eqref{Eq:Bound_cond_U_1} and~\eqref{Eq:Bound_cond_U_2}] for each problem in turn. Since charge neutrality is assumed to hold in bulk, we can also safely neglect the term $\propto \rho_{0}(r)\delta\phi(\mathbf{r})$ in Eq.~\eqref{Eq:Stokes_lin}.
Taking the curl of Eq.~\eqref{Eq:Force_free_Stokes} eliminates the pressure  and using the representation of the velocity field in terms of the poloidal Debye potential [Eqs.~\eqref{Eq:Velocity_field_sym}], yields for the scalar function $R=R(r)$ the ordinary linear differential equation
\begin{equation}
\mathscr{L}(\mathscr{L}(R))(r)=0,
\end{equation}
with the differential operator
\begin{equation}
\mathscr{L}=\frac{\textrm{d}^2}{\textrm{d}r^2}+\frac{2}{r}\frac{\textrm{d}}{\textrm{d}r}-\frac{2}{r^2}.
\end{equation}
An asymptotic solution then reads
\begin{align}\label{Eq:Asymptotic}
R(r) &\sim C_{N+1}+\frac{C_{N+2}}{r^2} \quad\text{for}\quad r\rightarrow\infty,
\end{align}
with constants  $C_{N+1}, C_{N+2}$  to be determined, where  the notation is  adopted
from Ref.~\cite{OBrien:JCS_74:1978}.  
By symmetry and linearity in $\mathbf{X}$ 
the perturbation in the scalar pressure field assumes the form  $\delta P(\mathbf{r}) = \pi(r) (\mathbf{r} /r) \cdot\mathbf{X}$ with a radially symmetric field $\pi(r)$ which can be calculated for large distances from Eq.~\eqref{Eq:Force_free_Stokes} to
\begin{equation}
\pi(r)\sim \eta \frac{\textrm{d}}{\textrm{d}r}\left(r\mathscr{L}(R )\right) = \frac{2 \eta C_{N+1}}{r^2} \quad\text{for}\quad r\rightarrow\infty.
\end{equation}
The magnitude of the force $\mathbf{F}=F \mathbf{X} / |\mathbf{X}|$ exerted by the fluid on the particle
\begin{equation}
F= -\int \textrm{d} S\left(\sigma_{rr}(\mathbf{r})\cos\vartheta-\sigma_{\vartheta r}(\mathbf{r})\sin\vartheta\right),
\end{equation}
is now evaluated from the viscous stresses in spherical coordinates \begin{subequations}
\begin{align}
\sigma_{rr}(\mathbf{r})&=- P(\mathbf{r}) + 2 \eta \frac{\partial u_r(\mathbf{r})}{\partial r} \nonumber  \\
&=- \eta \left( \frac{6 C_{N+1}}{r^2} + \frac{12 C_{N+2}}{r^4} \right) |\mathbf{X}| \cos\vartheta, \\
\sigma_{\vartheta r}(\mathbf{r})&= \eta \left[ \frac{1}{r}\frac{\partial u_r(\mathbf{r})}{\partial \vartheta} + \frac{\partial u_\vartheta(\mathbf{r})}{\partial r} - \frac{ u_\vartheta(\mathbf{r})}{r} \right]  \nonumber \\
&= - \eta \frac{6 C_{N+2}}{r^{4}}  | \mathbf{X}| \sin \vartheta.
 \end{align}
\end{subequations}
We thus arrive at
\begin{equation}
\mathbf{F}=8\pi\eta C_{N+1} \mathbf{X} ,
\end{equation}
and consequently it follows from Eq.~\eqref{Eq:D_T_force} that
\begin{equation}
D_{T}=\frac{C_{N+1}^{(2)}}{C_{N+1}^{(1)}} \frac{1}{T_0},
\end{equation}
where the constants $C^{(1)}_{N+1}, C^{(2)}_{N+1}$ have to be extracted from the asymptotic behavior of $R(r)$ [Eq.~\eqref{Eq:Asymptotic}] for problem (1) and (2). 
As an additional result, we obtain the diffusion coefficient for the particle as $D=k_{\text{B}} T_0/8\pi\eta C^{(1)}_{N+1}$.

\subsection{Nondimensional formulation and reference scales}\label{Sec:nondimensonal_problem}

We employ a dimensionless formulation, measuring lengths in units of the particle radius $a$ and electrostatic potentials in units of the thermal voltage $k_{\text{B}} T_0 /e$. The Poisson equation [Eq.~\eqref{Eq:Poisson_lin}] suggests then measuring surface charge densities in units of $\epsilon_0 \epsilon_r^0  k_{\text{B}} T_0/a e$, while  the viscosity in Stokes' equation [Eq~\eqref{Eq:Stokes_lin}] sets the unit of velocity to $U_0=\epsilon_0 \epsilon_r^0 (k_{\text{B}} T_0)^2 /e^2 \eta a$. 
Rather than using dimensionless concentrations $n_{i}(r) a^3$, we follow tradition and  introduce the dimensionless  concentrations by $n_{i}(r) /2 I$ (and similarly for the reference concentrations $n_{i,0}^b/2I$) with the constant ionic strength in the bulk solution
\begin{equation}
 I=\frac{1}{2}\sum_{i=1}^{N}z_{i}^{2}n_{i,0}^{\text{b}}.
\end{equation}
For a monovalent salt assuming completely dissociated ions, the dimensionless concentrations simplify to $n_{+,0}^{\text{b}}/2I=n_{-,0}^{\text{b}}/2I=1/2$ for cations ($+$) and anions ($-$) as the valences evaluate to $\pm 1$. Similar expressions can also be found for divalent or trivalent salts. Consequently, this renders the problem independent of the equilibrium ion bulk concentrations, except for the dimensionless inverse Debye screening length $\kappa_{0}$. It characterizes the limiting cases of a thin ($\kappa_{0}\gg 1$), respectively wide ($\kappa_{0}\ll 1$) double layer as compared to the particle radius $a$. Once we fix the dimension of the particle, $\kappa_{0}$ can only vary with the ionic strength $I$.

Then the Poisson-Boltzmann equation for the dimensionless equilibrium potential $\phi_{0}(r)$ reads 
\begin{equation}\label{Eq:Poisson_Boltzmann_scaled}
\frac{1}{r^{2}}\frac{\textrm{d}}{\textrm{d}r}\left[r^{2}\frac{\textrm{d}}{\textrm{d}r}\phi_{0}(r)\right]=-\kappa_{0}^{2}\sum_{i=1}^{N}z_{i}n_{i,0}^{\text{b}}\exp\left[-z_{i}\phi_{0}(r)\right],
\end{equation}
subject to the boundary conditions
\begin{subequations}\label{Eq:PB_scaled_BC}
\begin{align}
\lim\limits_{r\rightarrow\infty}{\phi_{0}(r)}&=0, \\
\frac{\textrm{d}\phi_{0}(r)}{\textrm{d}r}\bigg|_{r=1}&= - \sigma_0. \label{Eq:PB_scaled_BC_2}
\end{align}
\end{subequations}
Here $\sigma_{0}$ denotes the dimensionless bare colloidal surface potential.
Further, using the symmetry-adapted ansatz for the ionic potential 
 and the velocity field [Eqs.~\eqref{Eq:Velocity_field_sym} and ~\eqref{Eq:Debye_symmetry}], we obtain from Eqs.~\eqref{Eq:Omega_spelled_out} and~\eqref{Eq:Stokes_lin_decoupled} the coupled linear ODEs in dimensionless form 
\begin{subequations}\label{Eq:ODEs_scaled}
\begin{align}
 &\mathscr{L}\omega_{i}(r)-z_{i}\frac{\textrm{d}\phi_{0}(r)}{\textrm{d}r}\left[\frac{\textrm{d}\omega_{i}(r)}{\textrm{d}r} - \text{Pe}_{i} \frac{2 R(r)}{r}\right]= \nonumber \\
 &\qquad z_i\frac{\textrm{d} \phi_0(r)}{\textrm{d}r } \begin{cases}
 \text{Pe}_i, & (1) \\
  z_i \left[\phi_0(r)+ \phi^{\text{th}}\right]-1,  & (2) ,                                                                
 \end{cases} \\
 &\mathscr{L}(\mathscr{L} R)(r)+\kappa_0^{2}\frac{\textrm{d} \phi_0(r)}{\textrm{d} r}\sum_{i=1}^{N} z_{i}  n_{i}^{0}(r) \frac{\omega_{i}(r)}{r}=
\nonumber \\
&-\kappa_0^{2}\frac{\textrm{d} \phi_0(r)}{\textrm{d} r}\sum_{i=1}^{N} z_{i}  n_{i}^{0}(r)\left[\begin{cases}
				0, & (1) \\
				z_{i}\phi_{0}(r)+z_{i}\phi^{\text{th}}-S_{T}^{i}T_{0}, & (2)
				\end{cases}\right] \nonumber \\
&\qquad\qquad-\begin{dcases}
	0, & (1) \\
	\alpha\frac{\textrm{d}\phi_{0}(r)}{\textrm{d}r}\frac{\textrm{d}^{2}\phi_{0}(r)}{\textrm{d}r^{2}}, & (2) ,
	\end{dcases}
\end{align}
\end{subequations}
for the nondimensional functions $\omega_{i}(r)$ and $R(r)$. In the preceding equations, we have introduced the ionic P{\'e}clet number~\cite{Saville:ARFM_9:1977}  
\begin{equation}
\text{Pe}_{i}=\frac{U_{0}a}{D_{i}^{0}} ,
\end{equation}
quantifying the ratio between convective and diffusive ion transport.  The corresponding far-field boundary conditions translate to
\begin{subequations}\label{Eq:far_field_BC_scal}
\begin{align}
\lim\limits_{r\rightarrow\infty}{\omega_{i}(r)}&=0, \\
\lim\limits_{r\rightarrow\infty}{\frac{R(r)}{r}}&=0, \\
\lim\limits_{r\rightarrow\infty}{\frac{\textrm{d}R (r)}{\textrm{d}r}}&=0
\end{align}
\end{subequations}
and at the surface of the colloidal particle, the boundary conditions assume the form
\begin{subequations}\label{Eq:slipping_plane_BC_scal}
\begin{align}
&\frac{\textrm{d}\omega_{i}(r)}{\textrm{d}r}\bigg|_{r=1}=
\begin{cases}
     0, & (1) \\
    -z_{i}[\phi_{0}(r)|_{r=1}+ \phi^{\text{th}}], & (2),
\end{cases} \label{Eq:omega_scaled}\\
&\left. \frac{R(r)}{r} \right|_{r=1}=
\begin{cases}
     1/2, & (1) \\
      0, & (2),
\end{cases} \\
 &\left. \frac{\textrm{d}R(r)}{\textrm{d}r}
\right|_{r=1}-\left. \lambda\frac{\textrm{d}^{2}R(r)}{\textrm{d}r^{2}}\right|_{r=1}=
\begin{cases}
    1/2,  & (1) \\
     0,      & (2).
 \end{cases}         
\end{align}
\end{subequations}
Eventually, these equations are solved numerically to determine the thermal diffusion coefficient $D_{T}(\sigma, \kappa_{0}, \lambda)3T_{0}/2U_{0}a$ as a dimensionless function of the rescaled bare colloidal surface potential $\sigma_{0}$, the normalized inverse Debye width $\kappa_{0}$ and the reduced slip length $\lambda$ for different salt species. Similar to the electrophoresis problem~\cite{OBrien:JCS_74:1978}, the additional factor of $3/2$ is introduced for a convenient comparison with other theoretical approaches~\cite{Rasuli:PRL_101:2008, Burelbach:EPJE_42:2019}.

\section{Numerical solution of the differential equations}\label{Sec:Numerical_procedure}

In this section, we describe the numerical methods employed to obtain approximate solutions of the ODEs in dimensionless form as elaborated in the previous subsection for the relevant functions $\phi_{0}(r)$, $\omega_{i}(r)$ and $R(r)$. The Poisson-Boltzmann equation is solved relying on a Chebyshev spectral collocation method~\cite{Boyd:CFSM:2003, Canuto:SMFD:1988}. For the coupled linear ODEs [Eqs.~\eqref{Eq:ODEs_scaled}] a shooting method~\cite{Osborne:JMAA_27:1969}, together with asymptotic matching is applied adapting the solution procedure of O'Brien and White~\cite{OBrien:JCS_74:1978} for the electrophoresis problem.

\subsection{Solving the Poisson-Boltzmann equation with the Chebyshev spectral collocation method}\label{Sec:Chebyshev_collocation}

Since the dimensionless potential, as well as its first and second derivative are required as coefficients in Eqs.~\eqref{Eq:ODEs_scaled} and in the corresponding boundary condition [Eqs.~\eqref{Eq:omega_scaled}], we have to   determine numerically first these quantities from the nonlinear Poisson-Boltzmann equation.

Thus after mapping the half-infinite domain $[1, \infty)$  to the half-open interval $[-1, 1)$ by a diffeomorphism, this two-point boundary value problem (BVP) [Eqs.~\eqref{Eq:Poisson_Boltzmann_scaled} and~\eqref{Eq:PB_scaled_BC}] can be solved efficiently and with high accuracy by applying a Chebyshev spectral collocation method to the transformed BVP (see appendix \ref{App:Chebyshev_collocation}). Here, a nonlinear coordinate transformation of the form
\begin{equation}
\Phi_{L}: [-1, 1)\rightarrow [1, \infty), \quad t\mapsto r=L\frac{(1+t)}{(1-t)}+1
\end{equation}
is used, where $L>0$ denotes an adjustable mapping parameter. The advantage of the chosen algebraic transformation is its smoothness and robustness, i.e.\@ the decreased sensitivity on $L$~\cite{Boyd:JCP_45:1982,  Grosch:JCP_25:1977, Canuto:SMFD:1988}. 
Then, in the finite domain $[-1, 1]$ we approximate the solution to the problem $\phi_{0}(t)$ by a global Lagrange-interpolation polynomial of degree $M$~\cite{Canuto:SMFD:1988, Peyret:SMIVF:2002} that satisfies the mapped BVP at the  Chebyshev-Gauss-Lobatto points
\begin{equation}
t_{j}=\cos\left(\frac{j\pi}{M}\right), \quad j=0,\ldots, M.
\end{equation}
The $p$-th derivative ($p=1,2$) is obtained by differentiating the interpolant at these nodal points $\{t_{j}\}$, defining the discretized derivative operators which can be represented by Chebyshev differentiation matrices $\mathsf{D}^{(p)}$~\cite{Canuto:SMFD:1988, Boyd:CFSM:2003}. Accordingly, the numerical differentiation may be performed as 
\begin{equation}
\mathbf{y}^{(p)}=\mathsf{D}^{(p)}\mathbf{y},
\end{equation}
where $\mathbf{y}$ and $\mathbf{y}^{(p)}$ are the vectors of function values, respectively approximate derivative values at these nodes and $\mathsf{D}^{(p)}=\left(\mathsf{D}^{(1)}\right)^{p}$.
The transformed BVP is now converted to a set of $M+1$ nonlinear algebraic equations that are solved by the Newton-Raphson method with an appropriate initial guess (for further details see Appendix \ref{App:Chebyshev_collocation}). 
We choose the mapping parameter as equal to the dominant length scale of the solution $L=1/\kappa_{0}$, i.e. the Debye length in units of the particle radius  and vary the total number of Chebyshev nodes depending on the rescaled bare surface potential $\sigma_{0}$ ensuring a rapid convergence of the polynomial series coefficients for different $\kappa_{0}$. This rapidity facilitates the high accuracy of the calculated numerical solution, as well as the stability of the numerical scheme~\cite{Boyd:JCP_45:1982, Boyd:CMA_41:2001}.
Finally, an approximate solution for $\phi_{0}(r)$ is obtained on the unbounded interval $\left[1,\infty\right)$ in terms of a transformed barycentric interpolant~\cite{Berrut:SIAMR_46:2004} using the inverse transform $\Phi_{L}^{-1}$ (see Appendix \ref{App:Chebyshev_collocation}). Similar expressions for the first and second derivative of $\phi_{0}(r)$ are also derived.

\subsection{Solving the coupled linear ODEs with a shooting method and asymptotic matching}

The algorithm for solving the coupled set of linear ODEs [Eqs.~\eqref{Eq:ODEs_scaled}] is based on a predictor-corrector Adams-multistep method adaptively choosing both step size and order~\cite{Shampine:CSODE:1975}.
We start the numerical integration at large radial distance $r_{0}=1+20/\kappa_{0}$, i.e\@ in the bulk, with the asymptotic forms for the functions $\omega_{i}(r)$ and $R(r)$ and terminate it after reaching the rescaled (virtual) colloidal surface with $r=1$.
Neglecting  exponentially small terms due to the electrostatics in Eqs.~\eqref{Eq:ODEs_scaled} for $r\rightarrow\infty$, the asymptotic behavior can be obtained from
\begin{subequations}
\begin{align}
 \mathscr{L}\omega_{i}(r)=0, \\
 \mathscr{L}\left(\mathscr{L}R\right)(r)=0,
\end{align}
\end{subequations}
for both problems (1) and (2) obeying the far-field boundary conditions [Eqs.~\eqref{Eq:far_field_BC_scal}]. This yields
\begin{subequations}\label{Eq:asymp_form_scal}
\begin{align}
 \omega_{i}(r)&\sim\frac{C_{i}}{r^{2}}, \\
 R(r)&\sim C_{N+1}+\frac{C_{N+2}}{r^2},
\end{align}
\end{subequations}
for $r\rightarrow\infty$ with  asymptotic constants $C_{i}, \, i=1,\ldots, N+2$ for problems (1) and (2), respectively. The second expression is reminiscent of the results for the velocity field obtained in Sec.\ref{Sec:Force_calc}, however now in their nondimensional forms.
We are aiming to determine the set of asymptotic constants
\begin{equation}
 \mathbf{C}=\left(C_{1},\ldots,C_{N+2}\right)^{\text{T}},
\end{equation}
for the two problems  from the slipping plane boundary condition [Eqs.~\eqref{Eq:slipping_plane_BC_scal}]. The linearity of the coupled ODEs allows  writing a general solution as the following linear combination
\begin{equation}\label{Eq:lin_comb_sol}
 \mathbf{y}(r)=\mathbf{y}_{\text{part}}(r)+\sum_{k=1}^{N+2}C_{k}\mathbf{y}^{k}_{\text{hom}}(r),
\end{equation}
by superimposing a particular solution $\mathbf{y}_{\text{part}}(r)$ for each problem (1) and (2) with $N+2$ homogeneous solutions $\mathbf{y}_{\text{hom}}^{k}(r)$. Note that the homogeneous solutions $\mathbf{y}_{\text{hom}}^{k}(r)$ are the same for both problem (1) and (2). 

First, we define the $k$-th solution ($k=1,\ldots,N+2$) to the homogeneous problem as
\begin{equation}
 \mathbf{y}^{k}_{\text{hom}}(r)=\left(\omega_{1}^{k}(r),\ldots,\omega_{N}^{k}(r),R^{k}(r)\right)^{\text{T}}.
\end{equation}
In addition, the initial condition for this solution set is determined by the asymptotic forms [Eqs.~\eqref{Eq:asymp_form_scal}] in combination with the particular choice
\begin{equation}
 C_{i}^{k}=\delta_{ik},\quad i=1,\ldots,N+2 , 
\end{equation}
for the asymptotic constants $C_{i}$.
Utilizing these initial condition, we then solve for each value of $k=1,\ldots,N+2$ in turn the homogeneous forms of Eqs.~\eqref{Eq:ODEs_scaled}
\begin{subequations}
\begin{align}
 \mathscr{L}\omega_{i}(r)-z_{i}\frac{\textrm{d}\phi_{0}(r)}{\textrm{d}r}\left[\frac{\textrm{d}\omega_{i}(r)}{\textrm{d}r} - \text{Pe}_{i} \frac{2 R(r)}{r} \right]&=0,\\
\mathscr{L}(\mathscr{L} R)(r)+\kappa_0^{2}\frac{\textrm{d} \phi_0(r)}{\textrm{d} r} \sum_{i=1}^{N} z_{i}  n_{i}^{0}(r) \frac{\omega_{i}(r)}{r}&= 0, 
\end{align}
\end{subequations}
by numerical integration from $r=r_{0}$ down  to the virtual colloidal surface  at $r=1$. 

Second, to obtain a particular solution  denoted as
\begin{equation}
 \mathbf{y}_{\text{part}}(r)=\left(\omega_{1}(r),\cdots,\omega_{N}(r),R(r)\right)^{\text{T}},
\end{equation}
the inhomogenous ODEs [Eqs.~\eqref{Eq:ODEs_scaled}] are again numerically integrated from $r=r_{0}$ down to the slipping plane at $r=1$ for problems (1) and (2).
Here all asymptotic constants are set to zero,
\begin{equation}
 C_{i}=0 \quad  i=1,\cdots,N+2.
\end{equation}
Substituting the general solution [Eq.~\eqref{Eq:lin_comb_sol}] into the boundary conditions at the colloidal surface [Eqs.~\eqref{Eq:slipping_plane_BC_scal}], yields a linear system of $N+2$ simultaneous equations of the form
\begin{equation}\label{Eq:linear_problem_constants}
\mathsf{A} \cdot \mathbf{C}=\mathbf{B}
\end{equation}
for the $N+2$ asymptotic coefficients $\mathbf{C}$ for problems (1) and (2). The coefficient matrix $\mathsf{A}$ and the vector $\mathbf{B}$ for both problems can be found in Appendix \ref{App:Matrix_rep_asymp_sol}. We solve these equations by Gaussian elimination with maximum pivoting.
The method presented requires that the homogeneous ODEs have to be solved $N+2$ times and the inhomogeneous ODEs are solved once for each problem in turn. 

As the thermal diffusion coefficient is calculated from the asymptotic constants determined from the boundary condition at the slipping plane, our approach requires the functions to be resolved with high accuracy within the Debye double layer, as well as in the bulk region which may have considerably varying length scales. This is justified for the equilibrium potential $\phi_{0}(r)$, since the combination of the algebraic transformation, together with the Chebyshev collocation method yields high accuracy to possibly machine precision. Especially in the outer region (bulk), the transformation allows the potential to be sufficiently resolved notwithstanding that it decays exponentially. 

Furthermore, we also have extended the computational domain far enough to capture the power-law behavior of the respective functions. We have found by varying the radial distance $r_{0}$ that our choice $1+20/\kappa_{0}$ (corresponding to 20 Debye lengths) is an acceptable lower bound, balancing computational effort and accuracy of the results. The relative changes between each trial amounts to approximately $\approx 10^{-5}$ for all $\kappa_{0}$. Thus, the results for $D_{T}$ are identical within four to six (significant) digits.

\section{Results and Discussion}

In the following, we first validate the numerical procedure described in Sec.~\ref{Sec:Numerical_procedure} by comparing our results for the electrostatic potential and thermal diffusion coefficient with (semi-) analytical expressions from previous theoretical studies~\cite{Rasuli:PRL_101:2008, Debye:Phys_Z_24:1923}. 
Then, the theoretical work by Rasuli and Golestanian~\cite{Rasuli:PRL_101:2008} is carefully reexamined with the main focus on the effect of the thermoelectric field in bulk. 
Afterwards, a detailed comparison with a different theoretical approach~\cite{Burelbach:EPJE_42:2019} is performed, where besides the mentioned effect of electric migration in bulk also several other contributions to the thermal diffusion coefficient are investigated.
At the end, we compare experimental results obtained in Refs.~\cite{Reichl:PRL_112:2014, Duhr:PNAS_103:2006} on thermophoretic drift motion of single-stranded DNA, respectively polystyrene beads, to our theoretical predictions with particular emphasis on the hydrodynamic boundary condition, the effect of buffer dissociation and surface charging.
The characteristic parameters chosen to represent a typical aqueous electrolyte with different salt added, are summarized in Appendix~\ref{App:Relevant_parameter_values} and used to generate the Figs.~\ref{fig:plot_1}-\ref{fig:plot_6}. We point out, that all quantities in this section are presented in a nondimensional form (see Sec.~\ref{Sec:nondimensonal_problem} for the corresponding characteristic units) unless otherwise stated.

\subsection{Code validation in Debye-H{\"u}ckel approximation}

First, we test our numerical approach for the case of weakly charged colloids, where some analytic progress can be made.
The Debye-H{\"u}ckel approximation~\cite{Debye:Phys_Z_24:1923} for a weakly charged colloidal particle states that for $|z_{i}\phi_{0}|\ll 1$, the nonlinear Poisson-Boltzmann equation [Eq.~\eqref{Eq:Poisson_Boltzmann_scaled}] can be simplified by expanding the Boltzmann factor $\exp(-z_{i}\phi_{0})=1-z_{i}\phi_{0}+\mathcal{O}((z_{i}\phi_{0})^2)$ to obtain a linear differential equation for the rescaled equilibrium potential $\phi_{0}(r)$, using the electroneutrality condition in bulk. Then, assuming a monovalent salt an analytic solution for the potential and its first derivative fulfilling the boundary conditions [Eqs.~\eqref{Eq:PB_scaled_BC}] are readily obtained as
\begin{subequations}\label{Eq:Debye_Hueckel_analytic}
\begin{align}
\phi_{0}(r)&=\frac{\sigma_{0}}{1+\kappa_{0}}\frac{1}{r}\exp\left[-\kappa_{0}(r-1)\right], \\
\frac{\textrm{d}\phi_{0}(r)}{\textrm{d}r}&=-\left(\frac{1}{r}+\kappa_{0}\right)\phi_{0}(r).
\end{align}
\end{subequations}

\begin{figure*}[t]
\centering
\includegraphics[width=0.9\textwidth]{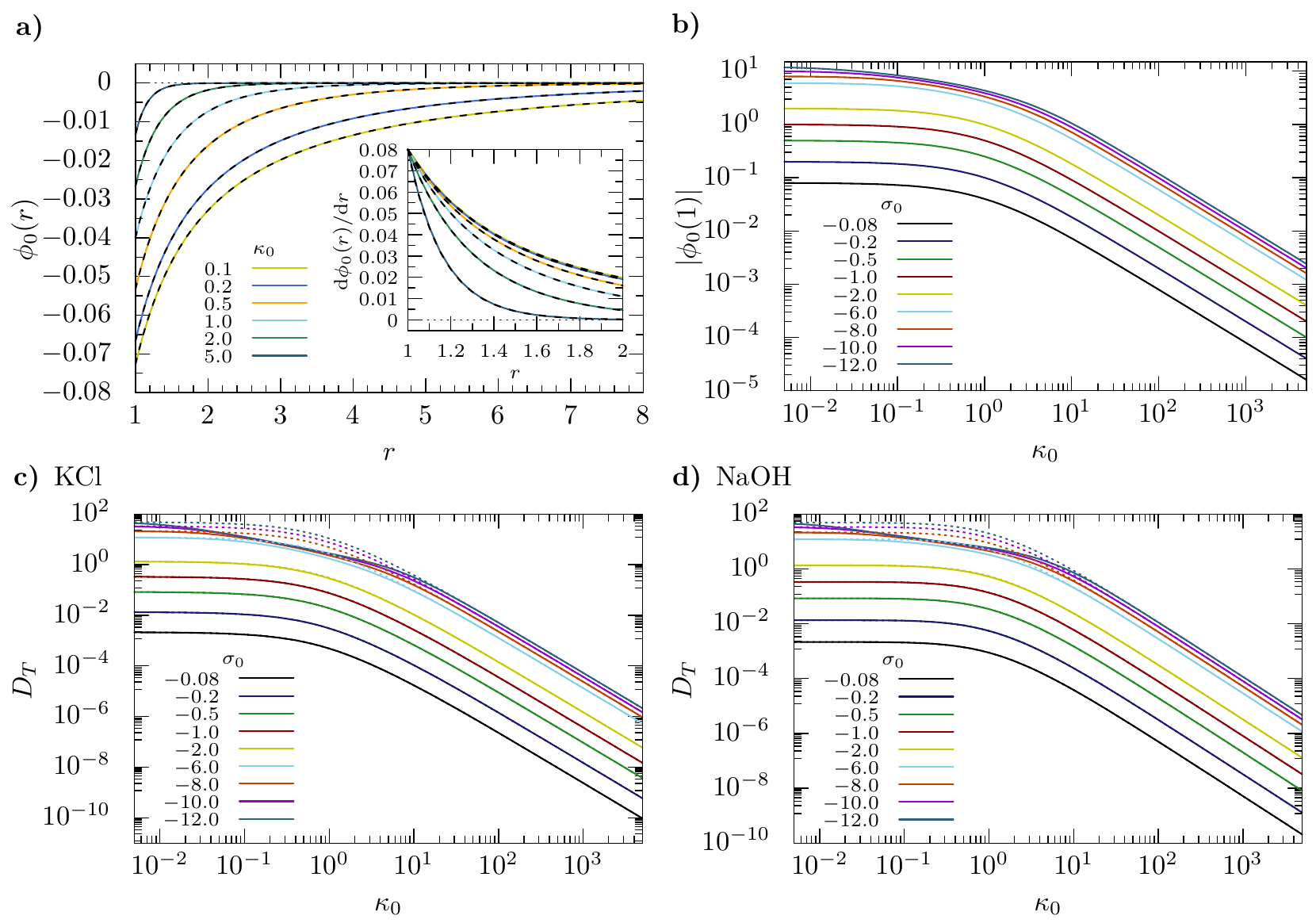}
\caption{a) Numerical results for the rescaled equilibrium potential and its first derivative (inset) as a function of the distance $r$ from the particle surface for intermediate Debye screening and weak negative surface charging, $\sigma_{0}=-0.08\, (\approx 2.1\,\text{mV})$. The black dashed lines denote the analytical results within the Debye-H{\"u}ckel approximation [Eqs.~\eqref{Eq:Debye_Hueckel_analytic}]. b) Variation of the numerically obtained zeta-potential values $\phi_{0}(1)$ with the dimensionless Debye screening length for different $\sigma_{0}$. Thermal diffusion coefficient for an aqueous solution with c) KCl and d) NaOH added. Compared are numerical solutions (solid lines) with analytic expressions from Ref.~\cite{Rasuli:PRL_101:2008} for a no-slip boundary condition.}
\label{fig:plot_2}
\end{figure*}

Our numerical approach to solve the nonlinear Poisson-Boltzmann equation by a Chebyshev collocation method (see Sec.~\ref{Sec:Chebyshev_collocation}) can now be validated by comparison with this analytic expression [Eqs.~\eqref{Eq:Debye_Hueckel_analytic}]. As shown in Fig.~\ref{fig:plot_2}a, for weak negative surface charging $\sigma_{0}=-0.08\, (\approx 2.1\,\text{mV})$ and intermediate Debye screening, our results are in perfect agreement with the theory.

Moreover, Rasuli and Golestanian~\cite{Rasuli:PRL_101:2008} have successfully derived semi-analytic formulas for the thermal diffusion coefficient by solving the coupled system of linear differential equations for the hydrodynamic solvent flow and the generalized ionic potentials  [Eqs.~\eqref{Eq:Omega_spelled_out} and~\eqref{Eq:Stokes_lin_decoupled}] within the Debye-H{\"u}ckel approximation.  In particular, the crossover between the two limiting cases of thin ($\kappa_{0}\gg 1$) and wide ($\kappa_{0}\ll 1$) Debye layers is elaborated. However, they have neglected the advection current and the coupling between ionic and electric potential functions which effectively disconnect the dynamics of the solutes from that of the solvent flow, providing an analytically tractable problem.  
We start by comparing our numerically determined results for the rescaled thermal diffusion coefficient $D_{T}$ with these analytic formulas for two aqueous solutions, adding exclusively the salt KCl, respectively NaOH for different bare surface potentials $\sigma_{0}$ and a no-slip boundary condition ($\lambda=0$). By artifically setting the thermoelectric potential to zero, we have modified our numerical treatment to account for the difference in the
ionic potential functions of both theoretical approaches (see also the next Sec.~\ref{Sec:Rasuli_model}). In addition, focusing on binary electrolytes equal ionic Soret coefficients $S_{T}^{+}=S_{T}^{-}=(\mathcal{S}_{T}^{+}+\mathcal{S}_{T}^{-})/2$ for cations $(+)$ and anions $(-)$ are used. Here, $\mathcal{S}_{T}^{\pm}=Q_{\pm}^{*}/k_{\text{B}}T_{0}^2$ is related to the ionic heat of transport due to water hydration effects for infinite dilution, see Ref.~\cite{Agar:JPC_93:1989} and Appendix~\ref{App:Soret_effect_bulk}. 
This helps in rearranging the pertinent equations into a form equivalent to those of  Ref.~\cite{Rasuli:PRL_101:2008}.
Then, for small bare surface potentials the numerical results agree very well with the predicted analytic expressions for the full range of Debye screening lengths and in fact, only for increasing bare potential values small deviations occur, since the Debye-H{\"u}ckel approximation ceases to be valid, as shown in Figs.~\ref{fig:plot_2}c and~\ref{fig:plot_2}d. Here the numerically calculated zeta-potential values $\phi_{0}(1)$ at the slipping plane, which varies with ionic strength and thus with the dimensionless Debye screening length $\kappa_{0}$, as $\sigma_{0}$ is fixed, corroborates this argument (see Fig.~\ref{fig:plot_2}b). The precise agreement with the semi-analytic formulas does not only confirm our numerical approach, but also shows that  the solution techniques of O'Brien and White~\cite{OBrien:JCS_74:1978} are reliably applicable to the problem of thermophoresis.

\subsection{Comparison with the model of Rasuli and Golestanian}\label{Sec:Rasuli_model}

\begin{figure}[htbp]
\centering
\includegraphics[width=0.48\textwidth]{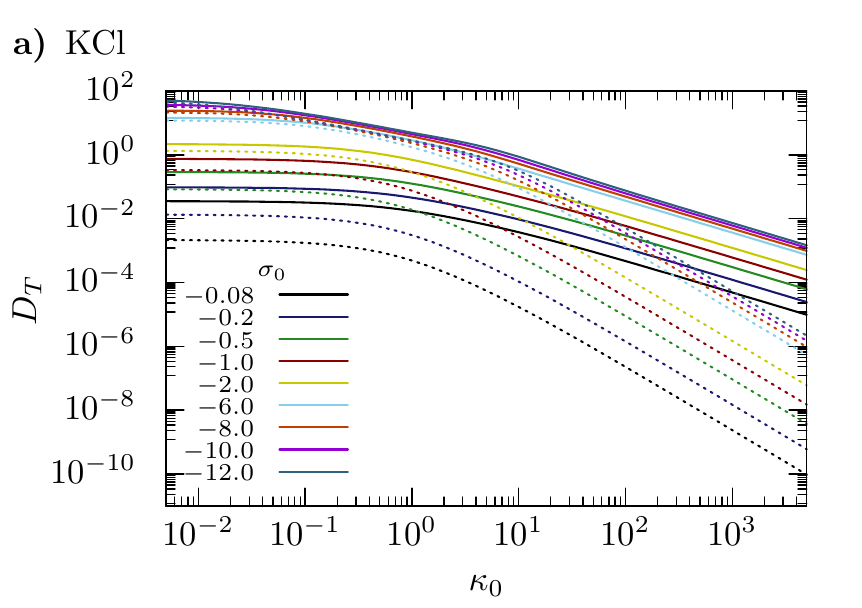}
\includegraphics[width=0.48\textwidth]{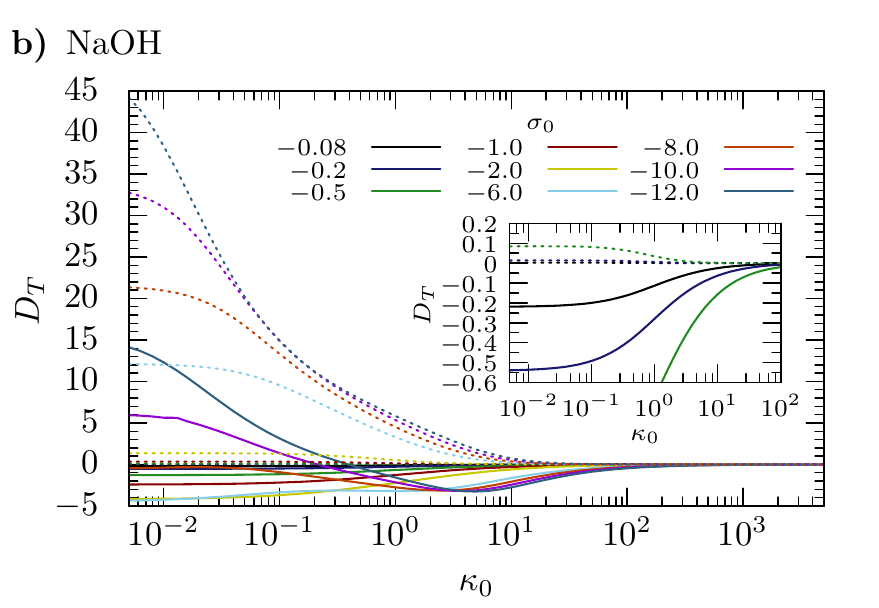}
\caption{Numerically obtained thermal diffusion coefficients for an aqueous solution in the presence of a) KCl and b) NaOH plotted against the inverse Debye screening length for different bare surface potential $\sigma_{0}$ and a no-slip boundary condition $\lambda=0$. The solid lines take into account the electrophoretic contribution to the ion hydration effect while the dashed lines discard it (as suggested in Ref.~\cite{Rasuli:PRL_101:2008}).}
\label{fig:plot_3}
\end{figure}

The theoretical continuum  model for thermophoresis provided in Ref.~\cite{Rasuli:PRL_101:2008} merely differs from our approach by the asymptotic behavior of the overall electrostatic potential. Since we properly account (to linear order) for the thermoelectric field behind the Debye double layer in bulk (see Eqs.~\eqref{Eq:bound_phi_full} and~\eqref{Eq:bound_conc_full} or Eqs.~\eqref{Eq:Boundary_pert_pot} and~\eqref{Eq:Boundary_pert_den}, respectively), Rasuli and Golestanian seemed to have implicitly discarded this electrophoretic contribution to the ion hydration effect in their treatment by not (directly) specifying a far-field boundary condition for the electrostatic potential. At least, it was not mentioned, neither in their paper~\cite{Rasuli:PRL_101:2008} nor in its Supplemental Material. Consequently, their choice of the ionic potential functions misses a (rescaled) term $\propto z_{i}\mathbf{r}\cdot\mathbf{E}^{\text{th}}=z_{i}\phi^{\text{th}}\mathbf{r}\cdot\nabla T$. In addition, they also have claimed, that an appropriate boundary condition for $r\rightarrow\infty$ consists of a vanishing potential functions $\omega_{i}(r)$ (see the Supplemental Material of Ref.\cite{Rasuli:PRL_101:2008}). Clearly within these assumptions, the steady-state distribution of the ionic solutes in bulk cannot be correctly recovered to linear order in thermal gradients (see Appendix~\ref{App:Soret_effect_bulk}) with ramifications for the thermal diffusion coefficient. 

Already for an aqueous solution titrated solely with KCl, which gives rise to a rather weak thermoelectric effect, respectively electrophoretic contribution with $\phi^{\text{th}}=-0.42$, deviations from our numerical results for the thermal diffusion coefficient over the whole range of inverse Debye screening lengths and for different surface potentials with a no-slip boundary condition become apparent (Fig.~\ref{fig:plot_3}a). Although their theory correctly predicts the sign of $D_{T}$, the difference increases up to six orders of magnitude once very thin double layers are considered. 
However, the discrepancies become even more prominent, when accounting for electrolytes with a strong thermoelectric effect. While for an aqueous solution adding exclusively the base NaOH ($\phi^{\text{th}}=-2.8$), the theoretical model of Rasuli and Golestanian~\cite{Rasuli:PRL_101:2008} yields only strictly positive thermal diffusion coefficients in the full parameter range, our numerical results for the transport coefficient $D_{T}$ with $\lambda=0$ show an inverse thermophoretic effect ($D_{T}<0$) for weak charging, together with a sign reversal around $\kappa_{0}\approx 1$, as the bare surface potential approaches large values (see Fig.~\ref{fig:plot_3}b). The work in Ref.~\cite{Burelbach:EPJE_42:2019} strongly supports our findings (see also Sec.~\ref{Sec:Burelbach} for details) rendering the ambiguous treatment of the boundary condition for the electrostatic potential and the corresponding choice of the ionic potential functions in Rasuli and Golestanian's 
work~\cite{Rasuli:PRL_101:2008} exclusively applicable in the limit of very small thermoelectric potentials $\phi^{\text{th}}\ll 1$. This severe restriction holds only for a few salt species, such as LiCl and NaF as the magnitude of the (rescaled) thermoelectric potential can reach up to $\approx 3$ ($\approx 100\,\text{mV}$) and its sign depends strongly on the relative difference of the ionic heat of transport, see Ref.~\cite{Wuerger:RPP_73:2010} and Appendix~\ref{App:Relevant_parameter_values}.
To account for this inconsistency, Rasuli and Golestanian also incorporated a possible salt dependence of $\mathcal{S}_{T}^{i}$ in their theoretical treatment when comparing with experiments which has not improved the situation yet. 

\subsection{Relation to the work of Burelbach and Stark}\label{Sec:Burelbach}

\begin{figure*}[t]
\centering
\includegraphics[width=0.76\textwidth]{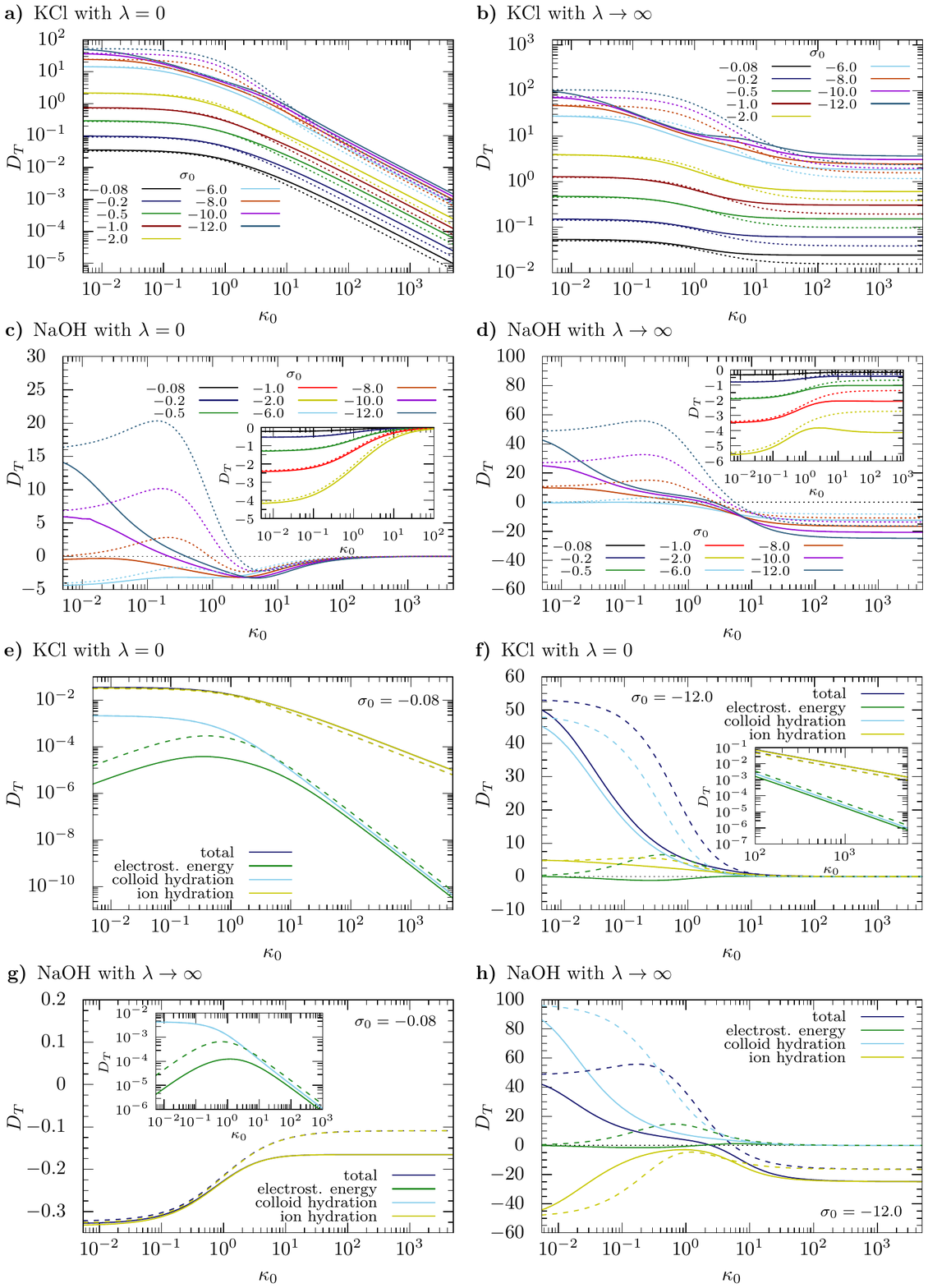}
\caption{Net thermal diffusion coefficient (a-d) and its different contributions (e-h) of a colloid with different bare surface potentials $\sigma_{0}$ for an aqueous electrolyte in the presence of the salt KCl, respectively NaOH as a function of the inverse Debye screening length for a no-slip ($\lambda=0$) and a perfect slip boundary ($\lambda\rightarrow\infty$). Our predictions and results from Ref.~\cite{Burelbach:EPJE_42:2019} are shown in solid and dashed-dotted lines, respectively.}
\label{fig:plot_4}
\end{figure*}

A different semi-analytical formula for the transport coefficient of a weakly charged colloidal particle with a hydrodynamic slipping surface undergoing thermophoresis has been proposed by Burelbach and Stark~\cite{Burelbach:EPJE_42:2019}.
Based on an alternative hydrodynamic approach~\cite{Burelbach:EPJE_41:2018} within the framework of non-equilibrium thermodynamics using Onsager's reciprocity relations, the colloidal drift velocity can be derived irrespective of how the screening length $1/\kappa_{0}$ compares to the particle size $a$. In our approach, we strongly rely on momentum conservation in a force-free system to obtain the thermal diffusion coefficient beyond the limiting cases of strong and weak screening.
Again, we numerically determine the thermal diffusion coefficient for a colloidal particle with different surface charging immersed in a water-based electrolyte solution with the salt KCl, respectively the base NaOH added. In Fig.~\ref{fig:plot_4}, the results are presented as a function of the inverse Debye screening length, together with the predictions from Ref.~\cite{Burelbach:EPJE_42:2019} for both, no-slip ($\lambda=0$) and perfect-slip ($\lambda\rightarrow\infty$) boundary condition. 

In general, their calculations for the rescaled thermophoretic mobility predict qualitatively similar behavior within the range of our testing parameters and both salts. Especially, the overall strong enhancement of $D_{T}$ in magnitude, together with the flattening out to a plateau for strong shielding $\kappa_{0}\gg 1$ as the slip length $\lambda$ is increased, are common features (Figs.~\ref{fig:plot_4}b and~\ref{fig:plot_4}d). In addition, for bare surface potentials $|\sigma_{0}|\gtrsim 6.0$, the sign reversal in the thermal diffusion coefficient for the base NaOH occurring independently of the slip length when $\kappa_{0}\approx 1$ is also covered by both theoretical models (Figs.~\ref{fig:plot_4}c and~\ref{fig:plot_4}d).
Nevertheless care has to be taken again, when comparing our predictions with those of Ref.~\cite{Burelbach:EPJE_42:2019}, since for an increasing $\sigma_{0}$, the dimensionless zeta potential $\phi_{0}(1)$ can become large (Fig.~\ref{fig:plot_2}b), such that the Debye-H{\"u}ckel approximation is no longer valid. Thus, the theoretical approach suggested in Ref.~\cite{Burelbach:EPJE_42:2019} does not apply in this regime. Fortunately, we do not encounter this problem, since the rescaled potential $\phi_{0}(r)$ is calculated from the full nonlinear Poisson-Boltzmann equation [Eq.~\eqref{Eq:Poisson_Boltzmann_scaled}]. Consequently, our findings suggest for screening lengths $\kappa_{0}\lesssim 1$ and large $\sigma_{0}$ a different though still complex behavior. To gain further insight into it, we also have computed the various contributions to the net thermophoretic transport coefficient $D_{T}$ as mentioned in Sec.~\ref{Sec:diff_contributions_thermophoresis} for different $\sigma_{0}$.  Here we display only both limiting cases of weak ($\sigma_{0}=-0.08$) and strong ($\sigma_{0}=-12.0$) charging for illustration purposes (see Figs.~\ref{fig:plot_4}e-h). Independent of the slip length $\lambda$ and strength of the thermoelectric effect, encoded in $\phi^{\text{th}}$, the term arising from the colloidal hydration is still the dominant contribution, yet significantly smaller as compared to the predictions from Ref.~\cite{Burelbach:EPJE_42:2019} (Figs.~\ref{fig:plot_4}f and~\ref{fig:plot_4}h). Consequently, our numerical results exhibit neither an extended shoulder in the curve for the salt KCl (see Fig.~\ref{fig:plot_4}a and~\ref{fig:plot_4}b) nor a pronounced peak in the function for the base NaOH (see Fig.~\ref{fig:plot_4}c and~\ref{fig:plot_4}d) around $\kappa_{0}\approx 1$.
In the limit of thin Debye double layers, respectively high ionic strength, Burelbach und Stark~\cite{Burelbach:EPJE_42:2019} have derived an analytic expression for the dimensionless thermal diffusion coefficient
\begin{equation}
D_{T}=\mu_{e}\phi^{\text{th}},
\end{equation}
as the product of the thermal potential and the electrophoretic mobility 
\begin{equation}
\mu_{e}=\begin{dcases}
						\frac{2\sigma_{0}}{3\kappa_{0}}, & \text{for $\lambda=0, \, \kappa_{0}\gg 1$} \\
						\frac{2}{3}\frac{\sigma_{0}\lambda}{(1+2\lambda)}, & \text{for $\lambda\neq 0, \, \kappa_{0}\ggg 1$}
			 \end{dcases}.					
\end{equation}
Hence, independent of the salt added $D_{T}$ converges either to zero for a no-slip boundary condition (Figs.~\ref{fig:plot_4}a and~\ref{fig:plot_4}c) or to a constant value, as the slip-length is increased (Figs.~\ref{fig:plot_4}b and~\ref{fig:plot_4}d). Whether a negative thermophoretic effect occurs, depends on the sign of $\sigma_{0}$ and $\phi^{\text{th}}$. Furthermore, the ion hydration effect is presumed to be the dominant contribution to the transport coefficient for strong screening and arbitrary slip length $\lambda$ (Figs.~\ref{fig:plot_4}e-h).
Our numerical predictions support all these findings, although we have observed a different scaling  for the thermophoretic transport coefficient $D_{T}$ and particularly its ionic hydration contribution since the electrophoretic mobility $\mu_{e}$ differs by a factor of $\approx 1.5$, as compared to the results of Ref.~\cite{Burelbach:EPJE_42:2019}, yielding the famous Helmholtz-Smoluchowski~\cite{Smoluchowski:BASC_182:1903} expression $\sigma_{0}/\kappa_{0}$ for $\lambda=0$ and a generalized version of it for a very thin Debye double-layer with a slipping boundary condition. In Ref.~\cite{Burelbach:EPJE_42:2019}, they also offer a possible explanation for this discrepancy referring to the dielectric permittivity of the colloid which was assumed to be equal in their treatment, whereas we have considered it to be negligible. Here, they have used a similar argument as in their treatment of the heat flow in the boundary layer approximation (see Appendix of Ref.~\cite{Burelbach:EPJE_41:2018}).

Nevertheless, for weak surface charging (Debye-H{\"u}ckel approximation) both predictions are in very good agreement with maximal relative deviations remaining below $5\%$ over a wide range of Debye screening lengths $\kappa_{0}$ (see Figs.~\ref{fig:plot_4}a and~\ref{fig:plot_4}b, together with the inset of Figs.~\ref{fig:plot_4}c and~\ref{fig:plot_4}d). In particular, at low ionic strength ($\kappa_{0}\ll 1$) both results seem to obey an identical limiting behavior, which also has been obtained in Ref.~\cite{Morthomas:EPJE_27:2008} within the point-particle limit $a\rightarrow 0$. Besides that, perfect accordance is also achieved for the colloidal hydration contribution to thermophoretic transport (Figs.~\ref{fig:plot_4}e and~\ref{fig:plot_4}g).

\subsection{Thermophoresis of single-stranded DNA}\label{Sec:Thermophoresis_DNA}

In this section, we compare our predictions for the thermophoretic transport coefficient to experimental results from Ref.~\cite{Reichl:PRL_112:2014} on 22mer single-stranded DNA molecules immersed in a TRIS-HCl (tris(hydroxymethyl)aminomethane-hydrochloride) buffered aqueous electrolyte with different monovalent salts added. The measurements have been conducted at room temperature and with 1 mM TRIS-HCl buffer to stabilize the pH-value around $7.5$.
These oligonucleotides exhibit a hydrodynamic radius of the order of the Debye length ($1/\kappa_{0}\approx 1$) and carry a rather high negative surface charge ($|\sigma_{0}|\lesssim 6$), requiring the electrical potential to be derived from the full nonlinear Poisson-Boltzmann equation [Eq.~\eqref{Eq:Poisson_Boltzmann_scaled}]. Thus, our theoretical approach provides a promising candidate to be tested against the experimental measurements.
In general, the effect of buffer dissociation on the thermophoretic transport coefficient has been ignored when fitting data points from experiments since the ionic heat of transport (see Appendix~\ref{App:Soret_effect_bulk}), as well as the ion mobilities of the buffer molecules are not known or difficult to obtain experimentally~\cite{Wuerger:PRL_101:2008, Rasuli:PRL_101:2008, Reichl:PRL_112:2014, Reichl:PRE_91:2015}. Yet, for the given pH-value, the TRIS-HCl buffer is almost fully dissociated and hence the contribution from the TRIS-$\text{H}^{+}$ cation to the thermoelectric effect cannot be neglected, as it sets a lower bound for $\kappa_{0}$ when the salt concentration is decreased. Although, the oxonium ($\text{H}_{3}\text{O}^{+}$) and hydroxide ions ($\text{OH}^{-}$) serve as a very efficient source for the thermoelectric potential, their influence can be safely ignored for the given pH-value~\cite{Reichl:PRL_112:2014}.

\begin{figure}[htbp]
\centering
\includegraphics[width=0.48\textwidth]{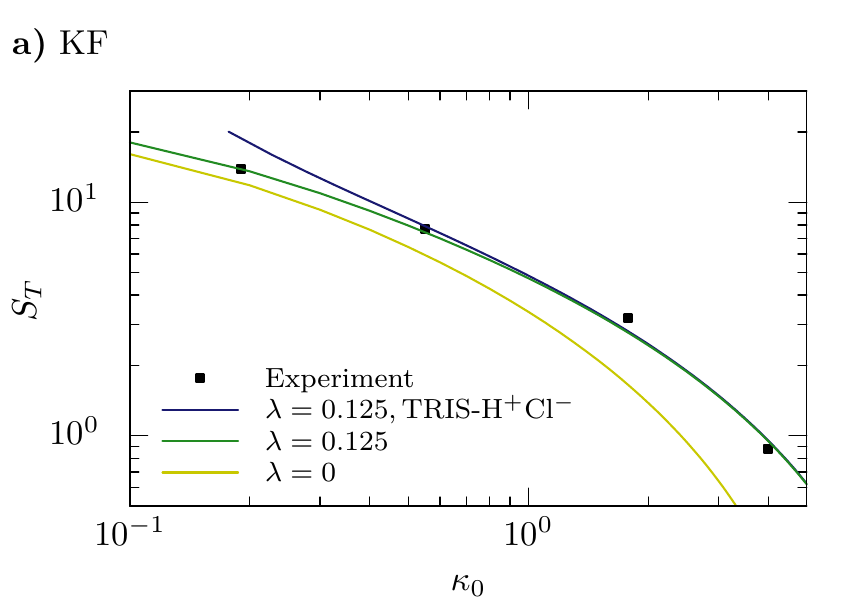}
\includegraphics[width=0.48\textwidth]{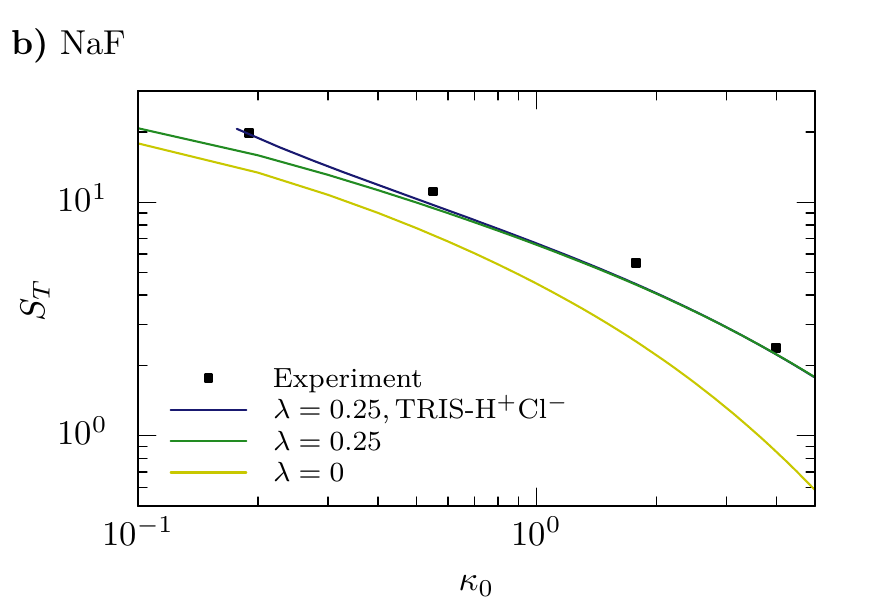}
\caption{Soret coefficient $S_{T}$ as a function of the inverse Debye screening length $\kappa_{0}$ for 22mer single-stranded DNA in the presence of a) KF and b) NaF. Symbols correspond to the experimental data from Ref.~\cite{Reichl:PRL_112:2014}, lines represent numerical predictions without any free fitting parameters for $a=1.7\,\text{nm}$, an effective charge of $Z=-13.8$ and different slip length $\lambda$. The theoretical calculations also take into account the influence of buffer dissociation.}
\label{fig:plot_5}
\end{figure}

In the presence of NaF and KF, the experimental data is well fitted by our numerical results for the dimensionless Soret coefficient $S_{T}$ without any free fitting parameters provided that a partial hydrodynamic slip is imposed at the DNA surface (see Fig.~\ref{fig:plot_5}a and~\ref{fig:plot_5}b). In particular, concerning the salt NaF a slip length of $\lambda=0.25$ ($\approx 0.43\,\text{nm}$) is used, whereas changing the cation, $\text{Na}^{+}\rightarrow\text{K}^{+}$ yields an even smaller value of $\lambda=0.125$ ($\approx 0.21\,\text{nm}$). 
For both salts a similar trend yet lower in magnitude is predicted for $\lambda=0$. 
Here, the Soret coefficient relates to the nondimensional thermal diffusion coefficient as follows~\cite{Burelbach:EPJE_42:2019}
\begin{equation}
S_{T}=D_{T}\frac{3a}{2\ell_{\text{B}}}\frac{1+2\lambda}{1+3\lambda},
\end{equation}
where $\ell_{\text{B}}=e^2/4\pi\epsilon_{0}\epsilon_{\text{r}}^{0}k_{\text{B}}T_{0}$ is defined as Bjerrum length, resulting from the balance of electrostratic and thermal energies. For water at room temperature, it takes the value $0.7\,\text{nm}$.
In addition, the fitting in Figs~\ref{fig:plot_5}a and~\ref{fig:plot_5}b has been carried out with a hydrodynamic radius of $a=1.7\,\text{nm}$ and an effective charge number $Z=-13.8$, connected to the bare surface potential by $\sigma_{0}=Z\ell_{\text{B}}/a$. Owing to the fact, that the average values $a=2\,\text{nm}$ and $Z=-11.6$ obtained in  Refs.~\cite{Reichl:PRL_112:2014, Reichl:PRE_91:2015} from experiments display rather big uncertainties, we have achieved a reasonable agreement with these numbers and consequently $a$, as well as $Z$ are not used as free-fitting parameters. Moreover, the effective charge per base pair $Z/22=0.63$ matches also with electrophoresis results using coarse-grained molecular-dynamics simulations~\cite{Reichl:PRL_112:2014, Grass:PRL_100:2008} and the value $\lambda=0.25$ for the slip length is consistent with current experiments in Ref.~\cite{Galla:NL_14:2014} on electrophoresis of DNA in nanopores, where a value of $\lambda=0.29$ have been suggested to explain their findings.
However, we can only speculate about the salt-dependent decrease in the hydrodynamic slip at the DNA surface. Obviously, modeling the single-stranded DNA molecule as a spherical particle with a homogeneously distributed surface charge neglects some of its important structural properties. The nucleo\-bases inside the DNA grooves are hydrophobic, leading to large hydrodynamic slip effects~\cite{Bocquet:SM_3:2007}, while the negatively charged phosphate groups of the backbone are known to be hydrophilic. However, the latest atomistic molecular dynamic simulation~\cite{Kesselheim:PRL_112:2014} provides evidence also for a non-zero tangential velocity along the DNA backbone. Possibly, the $\text{K}^{+}$ ions provides an enhanced efficiency in shielding these hydrophilic regions as the ionic radii of the cations $\text{K}^{+}$ and $\text{Na}^{+}$ differ by around $30\;\%$ and it is more likely for them to be located nearby the negatively charged phosphate groups due to electrostatic interactions, resulting in a smaller overall hydrodynamic slip length $\lambda$ for our simplified model. It is also likely, that a nonuniform surface conductivity~\cite{Mangelsdorf:JCSFT_86:1990, Carrique:JCIS_227:2000, Carrique:JCIS_243:2001, Khair:PF_21:2009}, which we did not account for in our theory, can effectively reduce the hydrodynamic slip. 

To incorporate the effect of buffer dissociation in the numerical calculations, we follow Ref.~\cite{Burelbach:EPJE_42:2019} and choose for the ionic heat of transport (see Appendix~\ref{App:Soret_effect_bulk}) of the TRIS-$\text{H}^{+}$-ion the same value as for the $\text{Na}^{+}$-ion.
In addition, the data for the mobility necessary to determine the ionic P{\'e}clet number (see Appendix~\ref{App:Relevant_parameter_values}) in the corresponding equations [Eqs.~\ref{Eq:ODEs_scaled}] is taken from a similar organic compound, the amino acid leucine~\cite{Wronski:JCA_657:1993}. The influence of the buffer dissociation is highlighted by changing the salt concentration and keeping the one of the buffer fixed. This is illustrated in Figs.~\ref{fig:plot_5}a and~\ref{fig:plot_5}b for the different salts KF and NaF.
While for intermediate Debye lengths ($1/\kappa_{0}\approx 1$), the contribution from the dissociated buffer ions to the Soret coefficient is of little significance independent of the added salt, in the regime of low ionic strength ($\kappa_{0}\lesssim 0.5$) it is to a large extent determined by the buffer ions which only moderately improve the agreement with the experimental data, especially for the salt NaF (see Fig.~\ref{fig:plot_5}b).
In general, our findings are in accord with the results obtained in Ref.~\cite{Burelbach:EPJE_42:2019}, except for the decrease in the hydrodynamic slip length as the cations are exchanged.

When accounting for the buffer molecules, the electrolyte consists of two monovalent salts which are assumed to be fully dissociated. Hence, both buffer ions $\text{TRIS-H}^{+}$ and $\text{Cl}^{-}$, together with the ions for the different salts KF and NaF are present in the aqueous solution with corresponding valences $\pm 1$.  Consequently, the dimensionless concentrations (see Sec.~\ref{Sec:nondimensonal_problem}) no longer evaluate to a constant, rather they explicitly depend on the Debye screening length $1/\kappa_{0}$ via 
\begin{subequations}\label{Eq:Dim_concentrations_buffer}
\begin{align}
\frac{n_{\text{buf}}}{2I}&=\frac{1}{2}\left(\frac{\kappa_{0}^{\text{buf}}}{\kappa_{0}}\right)^2, \\
\frac{n_{\text{s}}}{2I}&=\frac{1}{2}\left[1-\left(\frac{\kappa_{0}^{\text{buf}}}{\kappa_{0}}\right)^2\right],
\end{align}
\end{subequations}
where we have defined $n_{\text{s}}$ as the dimensional equilibrium bulk concentration of the added salt ions and $n_{\text{buf}}$ as the respective concentration of the buffer ions. Here $\kappa_{0}^{\text{buf}}=(8a^2 \pi\ell_{\text{B}}n_{\text{buf}})^{1/2}$ represents the dimensionless number for the inverse Debye screening length $\kappa_{0}$ in the absence of salt. Then, by varying only $n_{\text{s}}$ a fixed buffer concentration of $n_{\text{buf}}=1\,\text{mM}$ yields a lower bound of $0.18$ for $\kappa_{0}$. Using now Eqs.~\eqref{Eq:Dim_concentrations_buffer}, the thermal potential can be recast into
\begin{align}
\phi^{\text{th}}=&-\frac{T_{0}}{2}\left[\left(\frac{\kappa_{0}^{\text{buf}}}{\kappa_{0}}\right)^{2}\left(\mathcal{S}_{T}^{\text{TRIS-H}^{+}}-\mathcal{S}_{T}^{\text{Cl}^{-}}\right)\right. \nonumber \\
&\left. +\left(1-\left(\frac{\kappa_{0}^{\text{buf}}}{\kappa_{0}}\right)^{2}\right)\left(\mathcal{S}_{T}^{+}-\mathcal{S}_{T}^{-}\right) \right],
\end{align} 
with $\mathcal{S}_{T}^{\pm}=Q_{\pm}^{*}/k_{\text{B}}T_{0}^2$ for cations $(+)$ and anions $(-)$ and corresponding values for the buffer molecules. Similar expressions can be derived for $n_{i}^{0}(r)/2I$ by applying the same strategy. Thus, these quantities, especially the thermal potential are essentially dominated by the ions of the dissociated buffer at low ionic strength. Apparently, the dependence on the Debye screening length $1/\kappa_{0}$ vanishes upon the presence of only one species of salt and we recover the case of a binary electrolyte.

\subsection{Thermophoretic motion of polystyrene beads}
\label{Sec:PSB}

We also carry out numerical calculations for the experiment of Duhr and Braun~\cite{Duhr:PNAS_103:2006} performed on carboxylate-modified polystyrene beads (PSBs) of various sizes in the Debye-H{\"u}ckel regime. Similar to the 22mer single-stranded DNA molecules,
these PSBs are immersed in an aqueous solution buffered with $n_{\text{buf}}=0.5\,\text{mM}$ TRIS-HCl at a pH-value of $7.6$ and are titrated solely with KCl at different concentrations. From free-flow electrophoresis measurements on PSBs with radius $a=40\,\text{nm}$ and identical carboxyl-surface modifications at fixed nondimensional Debye length $1/\kappa_{0}=0.24$, an effective surface charge density of $\sigma_{\text{el}}= -4\,500\,e/\text{\textmu{}m}^{2}$ has been observed. Thus, the colloidal bare surface potential takes different values $\sigma_{0}=-4\pi\sigma_{\text{el}}\ell_{\text{B}}/ea$ depending on the size of the PSBs.

\begin{figure}[htbp]
\centering
\includegraphics[width=0.5\textwidth]{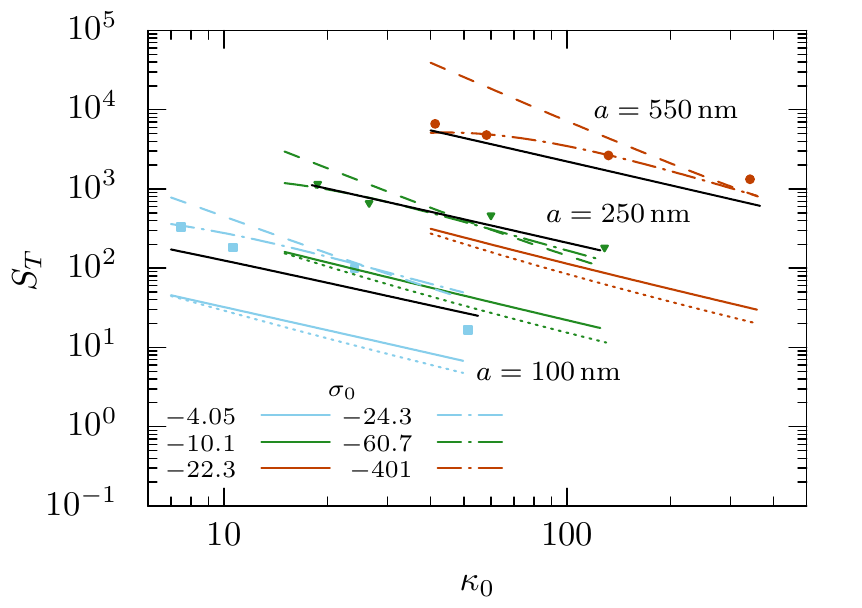}
\caption{Soret coefficient $S_{T}$ vs the inverse Debye screening length $\kappa_{0}$ for carboxyl modified polystyrene beads of radius $a=100, 250$ and $550\,\text{nm}$ exclusively titrated with KCl. Symbols relate to the experimental data from Ref.~\cite{Duhr:PNAS_103:2006}. Solid colored lines represent numerical predictions for a no-slip boundary condition $\lambda=0$ together with an effective surface charge density of $\sigma_{\text{el}}=-4\,500\,e/\text{\textmu{}m}^{2}$ corresponding to different bare surface potentials $\sigma_{0}$ and solid black lines are the analytic solutions from Ref.~\cite{Dhont:Langmuir_23:2007} for the same $\sigma_{\text{el}}$.
Dashed-dotted lines denote numerical results with artificially increased $\sigma_{0}$. Short and long-dashed lines represent the semi-analytical solutions from Ref.~\cite{Burelbach:EPJE_42:2019} for identical parameters.
}
\label{fig:plot_6}
\end{figure}

Then, a comparison between our theoretical predictions and the experimental measurements for the Soret coefficient $S_{T}$ is presented in Fig.~\ref{fig:plot_6} for three different PSB sizes and a no-slip boundary condition $\lambda=0$, since these PSBs hardly show a hydrodynamic slippage at their surface. In addition, no further adjustable fitting parameters are involved in the calculations. 
As a result, deviations between one and two orders of magnitude from our numerical results emerge. However, a satisfactory agreement can only be achieved by artificially increasing the surface charge density by a factor of $6$ for the PSBs with radii $a=100\,\text{nm}$ and $a=250\,\text{nm}$, while for the largest PSBs, it has to become $18$ times larger than the experimental determined value, which is far from every physical realistic number for colloidal charging. A very similar behavior is observed when applying the semi-analytical formula proposed by Burelbach and Stark~\cite{Burelbach:EPJE_42:2019} to calculate the Soret coefficient of a weakly charged colloidal particle (see Fig.~\ref{fig:plot_6}). Also for this theoretical approach, only an increase in the surface charge density leads to a good fit of the experimental data. 
In all these considerations, we have ignored contributions arising from buffer dissociation, as the qualitative behavior of the Soret coefficient of the PSBs changes only marginally.

\begin{figure}[htbp]
\centering
\includegraphics[width=0.48\textwidth]{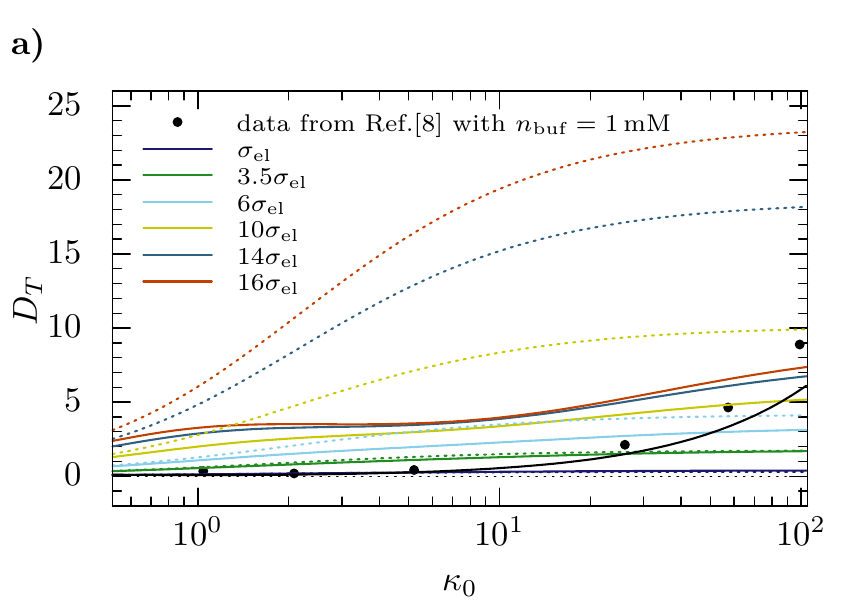}
\includegraphics[width=0.48\textwidth]{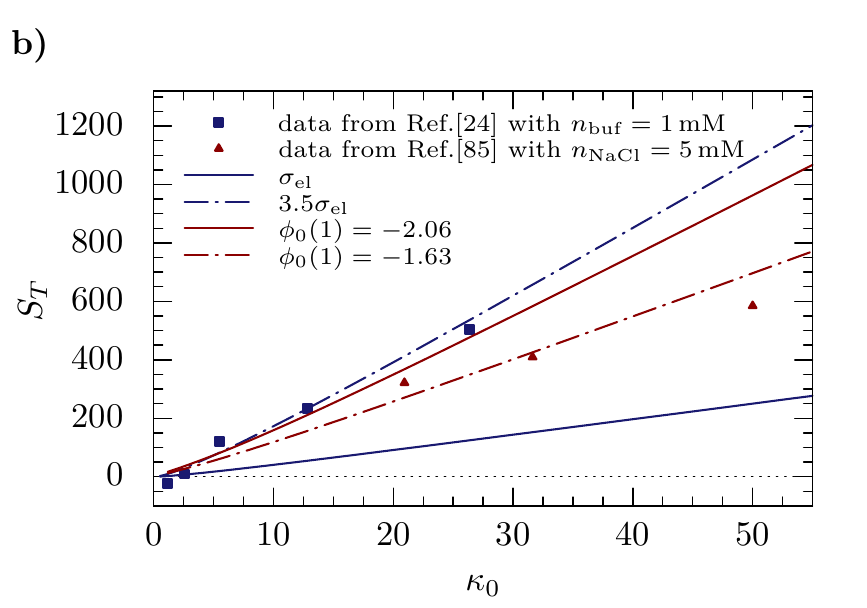}
\caption{a) Thermal diffusion coefficient $D_T$ and b) Soret coefficient $S_T$ vs the inverse Debye screening length $\kappa_{0}$ for polystyrene beads in an aqueous solution at room temperature with a fixed salt concentration corresponding to a varying particle radius. Measurement data (symbols) are taken from Refs.~\cite{Duhr:PNAS_103:2006, Braibanti:PRL_100:2008, Eslahian:Soft_Matter_10:2014}. 
a) Solid colored lines denote numerical predictions for a no-slip boundary condition and different surface charge densities $\sigma_{\text{el}}$. Dashed lines represent the semi-analytical results from Ref.~\cite{Burelbach:EPJE_42:2019} for different $\sigma_{\text{el}}$, the solid black line is the analytic solution from Ref.~\cite{Dhont:Langmuir_23:2007} for $\sigma_{\text{el}}=-4\,500\,e/\text{\textmu{}m}^{2}$.
b) Solid colored lines denote numerical predictions  for a no-slip boundary condition using the measured  surface charge densities $\sigma_{\text{el}}$, respectively zeta potentials $\phi_{0}(1)$.
Dashed-dotted lines correspond to numerical predictions using the surface charge density resp. the zeta potential as free parameter. 
 }
\label{fig:plot_7}
\end{figure}

Moreover, it is instructive to study the dependence of both, the thermal diffusion coefficient $D_{T}$ and the Soret coefficient $S_{T}$ on the size of the colloidal particle. Therfore, we compare our numerical predictions with experimental measurements conducted by Eslahian \emph{et al}~\cite{Eslahian:Soft_Matter_10:2014}, Braibanti \emph{et al}~\cite{Braibanti:PRL_100:2008}, and Duhr and Braun~\cite{Duhr:PNAS_103:2006} for PSBs with different surface modifications and stabilizing buffers. While the first experiment is performed on sulfated PSBs immersed in a deionized-water-based electrolyte adding $n_{\text{NaCl}}=5\,\text{mM}$ of the the salt NaCl, the last two experimental studies are carried out on carboxylated PSBs in an aqueous solution only buffered with $n_{\text{buf}}=1\,\text{mM}$ TRIS-HCl.
Our theoretical results for these quantities are shown in Fig.~\ref{fig:plot_7} as a function of the inverse Debye screening length. Here $\kappa_{0}$ varies exclusively with the particle radius $a$, since both the salt and buffer concentrations have been fixed in the experiments as well as for the numerical calculations, where we account for the relevant parameters of the dissociated buffer ions according to Sec.~\ref{Sec:Thermophoresis_DNA}.
Furthermore, the bare surface potential $\sigma_{0}$ becomes then also a function of the particle radius. 
A good agreement with the data from Ref.~\cite{Duhr:PNAS_103:2006} for the measured effective surface charge density $\sigma_{\text{el}}=-4\,500\,e/\text{\textmu{}m}^{2}$ can only be found for the smallest particle radii $a\lesssim 50\,\text{nm}$ and a drastic increase in $\sigma_{\text{el}}$ does not significantly improve the situation (see Fig.~\ref{fig:plot_7}a). This descrepancy is even more pronounced considering the semi-analytical predictions for the thermal diffusion coefficient derived by Burelbach and Stark~\cite{Burelbach:EPJE_42:2019} for equal surface charge density variations (see again Fig.~\ref{fig:plot_7}a).

The agreement is not much better, when comparing the predictions from theory with the experimental measurements from Braibanti \emph{et al}~\cite{Braibanti:PRL_100:2008} and Eslahian \emph{et al}~\cite{Eslahian:Soft_Matter_10:2014} for the Soret coefficient. Since both experiments are performed for various temperatures, we have extracted data at room temperature $T_{0}\approx 300\,\text{K}$. 
Assuming the same surface charge density $\sigma_{\text{el}}$ as in Ref.~\cite{Duhr:PNAS_103:2006} significant deviations are found for the experimental observations from Ref.~\cite{Braibanti:PRL_100:2008}. However, increasing its magnitude by a factor of 3.5 agreement can be obtained (see Fig~\ref{fig:plot_7}b). 
Similarly, relying on the measured zeta-potential for the data of Ref.~\cite{Eslahian:Soft_Matter_10:2014} overestimates the Soret coefficient.   
Yet, using a zeta potential roughly 20\% smaller than compared to the measured value yields reasonable agreement with our
theoretical approach (see again Fig.~\ref{fig:plot_7}b).  
Note, that we account for a constant zeta potential in the thermophoresis problem by replacing the boundary condition for the derivative of the dimensionless equilibrium potential [Eq.~\eqref{Eq:PB_scaled_BC_2}] at the colloidal surface by a constant surface-potential value $\phi_{0}(1)$. In the spectral collocation method this transforms then to the even simpler expression $\varphi_{M}=\phi_{0}(1)$ for the noncollocated endpoint $t_{M}$ (see Appendix~\ref{App:Chebyshev_collocation}) leading now to a set of $M$ nonlinear algebraic equations to be solved.

In contrast, other theoretical approaches derived within irreversible thermodynamics~\cite{Duhr:PNAS_103:2006, Dhont:Langmuir_23:2007, Kocherginsky:JCP_154:2021} under local thermodynamic equilibrium conditions are able to explain the experimental observations obtained by Duhr and Braun~\cite{Duhr:PNAS_103:2006} (see again Fig.~\ref{fig:plot_6} and ~\ref{fig:plot_7}a), while our model predictions can only fit the data from different experimental studies on PSBs in aqueous solutions~\cite{Duhr:PNAS_103:2006, Braibanti:PRL_100:2008, Eslahian:Soft_Matter_10:2014} when the parameters are strongly tuned such that the fitted surface charging differs strongly from the directly measured values. It appears questionable that these discrepancies can be rationalized by experimental uncertainties.  
This suggests, that other effects than these studied here should become important in controlling the behavior of the system.

\section{Summary and Conclusion}\label{Sec:Summary_conclusion}

In this work, we have numerically investigated the thermophoretic transport of a single spherical particle immersed in an electrolyte solution in linear response to an externally applied temperature gradient, addressing both moderately and highly charged solid surfaces exhibiting hydrodynamic slip for arbitrary Debye layer width. 
As a result of the linearization with respect to the spherically symmetric reference system at thermal equilibrium and by observing the axial symmetry of the thermophoresis problem regarding the imposed perturbations by the thermal gradient, a set of coupled ordinary differential equations has been systematically derived. Moreover, the dynamics of ions in the essentially electroneutral bulk solution is incorporated to linear order by appropriate far-field boundary conditions.  

In solving these linear differential equations, we have successfully utilized the solution techniques of O'Brien and White from their original treatment of the electrophoresis problem~\cite{OBrien:JCS_74:1978}.
The excellent agreement with (semi-) analytic expressions from former theoretical work~\cite{Rasuli:PRL_101:2008, Debye:Phys_Z_24:1923} for weakly charged particles provides confidence in our numerical calculations, thereby validating our numerical predictions of the thermal diffusion coefficient, as well as the electrostatic potential.

Moreover, in Ref.~\cite{Rasuli:PRL_101:2008} a similar theoretical model for thermophoretic motion has been presented, yet with a different treatment of the bulk solution behavior.
Consequently, we have examined their far-field boundary conditions for the potential functions by considering electrolyte solutions with both a strong and a weak thermoelectric effect. Our analysis reveals that the electrophoretic contribution to the colloidal hydration term is crucial to correctly predict the overall trend of the thermal diffusion coefficient. In particular, the inverse thermophoretic effect ($D_{T}<0$) for strong thermoelectric potentials cannot be explained as long as this term is missing. 

Only recently a description of colloidal thermophoresis based on Onsager's reciprocal relations has been introduced~\cite{Burelbach:EPJE_41:2018} and later on general expressions for the thermal diffusion coefficient of a weakly charged spherical particle in an aqueous electrolyte have been derived~\cite{Burelbach:EPJE_42:2019}. Altogether, our numerical predictions have essential features in common with their results. First, we also have observed the thermal diffusion coefficient to be sensitive to the hydrodynamic slip at the particle surface. In particular, this is accompanied by a constant thermal diffusion coefficient for strong shielding $\kappa_{0}\gg 1$ and non-vanishing slip length $\lambda\neq 0$, which is shown to be proportional to the electrophoretic mobility~\cite{Burelbach:EPJE_42:2019}. Second, for the base NaOH both models display a sign reversal in the thermophoretic transport coefficient independent of the slip length. This agreement corroborates our critical analysis of the far-field boundary condition for the electrostatic potential in Ref.~\cite{Rasuli:PRL_101:2008}.
In addition, we expect the negative thermophoretic effect to be an intrinsic characteristic of electrolytes with a strong thermoelectric potential, especially when hydroxide or oxonium ions are present, for example, the sodium hydroxide (NaOH) or hydrochloric acid (HCl). This numerical findings are also confirmed by experimental measurements on micellar solutions of sodium dodecyl sulfate~\cite{Vigolo:Langmuir_26:2010}.

Also, our numerical predictions agree well with the experimental data on 22mer single stranded DNA molecules in a TRIS-HCl buffered electrolyte~\cite{Reichl:PRL_112:2014}, which suggest the occurrence of hydrodynamic slippage along the surface of the DNA in accordance with Refs.~\cite{Burelbach:EPJE_42:2019, Galla:NL_14:2014}.
As part of this comparison, we further have probed the influence of the buffer dissociation on the thermal diffusion coefficient and ascertain, that for low overall ionic strength the buffer ions dominate the bulk behavior by setting the value for the thermoelectric potential.

Further, we have expected, that after modifying the theoretical model provided in Ref.~\cite{Rasuli:PRL_101:2008} by accounting for the dominant thermoelectric effect in bulk, our numerical results should also explain the experimental measurements on PSBs in the Debye-H{\"u}ckel regime~\cite{Duhr:PNAS_103:2006}. Unfortunately, only an unphysically large increase in the bare surface potential, respectively surface charge density leads to a sufficient agreement. 
By examining the dependence of our hydrodynamic approach for thermophoretic transport on the colloidal particle dimensions, we have revealed similar results. A varying surface charge density does not yield agreement between the theoretical predictions for the thermal diffusion coefficient and the experimental data measured by Duhr and Braun~\cite{Duhr:PNAS_103:2006} on PSBs and for the experiments conducted by Braibanti \emph{et al}~\cite{Braibanti:PRL_100:2008}, as well as Eslahian \emph{et al}~\cite{Eslahian:Soft_Matter_10:2014} we can achieve a consistent description only by tuning parameters to regimes which are hard to reconcile with the measured values.
In particular, this discrepancy in our theoretical analysis of the thermal diffusion coefficient of PSBs revives a prolonged debate initialized in Ref.~\cite{Astumian:PNAS_104:2007}. It deals with the question whether different regimes exist, where either the system is in local thermodynamic equilibrium by maximizing the number of microstates of the counter ions in the Debye layer surrounding the particle~\cite{Duhr:PRL_96:2006} or dissipation via local fluid flow dominates the phoretic motion thereby characterizing non-equilibrium transport. In the first regime, thermal fluctuations may become important, while in the other regime hydrodynamic stresses determine the phoretic drift velocity. The experiments on PSBs appear to fall into the first regime, where theoretical models based on irreversible thermodynamics are suitable and the hydrodynamic approach alone fail to account since it display only small corrections to the thermal diffusion 
coefficient~\cite{Kocherginsky:JCP_154:2021}. A detailed analysis of thermophoresis beyond thermodynamic equilibrium is provided in the companion paper~\cite{Mayer:PRL_130:2023}.

More generally speaking, our theoretical treatment corroborates the hydrodynamic character of thermophoretic motion to be a force-free interfacial phenomenon with local solvent flow in the vicinity of the colloid by showing that an explicit dependence on the hydrodynamic boundary condition occurs. This was also argued in Ref.~\cite{Burelbach:EPJE_42:2019}. Thus, a careful treatment of the surface properties of the colloidal particle plays a critical role in thermophoretic phenomena.
Moreover we have also generalized the force-free argument beyond the boundary layer approximation used in other theoretical approaches~\cite{Wuerger:PRL_101:2008, Wuerger:RPP_73:2010, Morthomas:EPJE_27:2008}.

\begin{acknowledgements}
We are grateful to  Bernhard Altaner for numerous helpful discussions. This work has been supported by the Austrian Science Fund (FWF): I5257-N.
\end{acknowledgements}

\appendix
\section{Soret effect of the ions}\label{App:Soret_effect_bulk}
 
We follow a commonly used approach describing ionic thermophoresis caused by hydration effects~\cite{Wuerger:RPP_73:2010, Guthrie:JCP_17:1949, Burelbach:EPJE_42:2019}. Here, the different ionic species are understood as a dilute gas of non-interacting charged particles enclosed by hydration layers of water molecules. The current densities of the ionic solutes
\begin{equation}\label{Eq:bulk_current}
 \mathbf{j}_{i}(\mathbf{r})=-D_{i}n_{i}^{b}(\mathbf{r})\left[\nabla \log n_{i}^{b}(\mathbf{r})+ \frac{Q_{i}^{*}}{k_{\text{B}} T} \frac{\nabla T}{T} -\frac{ z_{i}e\mathbf{E}^{\text{th}}(\mathbf{r})}{k_{\text{B}} T} \right],
\end{equation}
in the bulk solution with ion concentrations $n_{i}^{\text{b}}(\mathbf{r})$ comprise mass and thermal diffusion as well as thermoelectric migration. Here, the Einstein diffusion coefficient is evaluated at the reference temperature $T$ and $Q_{i}^{*}$ denotes the temperature-independent heat of transport for each ionic solute due to hydration by surrounding water molecules in the limit of infinite dilution~\cite{Helfand:JCP_32:1960, Agar:JPC_93:1989, Takeyama:JSC_17:1988}. 

Switching on the temperature gradient, the corresponding currents [Eq.~\eqref{Eq:bulk_current}] accumulate ions in a thin layer of thickness $\sim 1/\kappa_{0}$ close to the hot and cold boundaries of the system. Then the   thermoelectric field $\mathbf{E}^{\text{th}}(\mathbf{r})$ is fixed by the steady state of the solutes, where the ion currents $\mathbf{j}_{i}(\mathbf{r})=0$ vanish. This may be justified by the significantly slower reaction of the colloidal particle as compared to the ions~\cite{Burelbach:EPJE_41:2018, Dietzel:JFM_813:2017}. In  bulk, we further use the condition of  local charge neutrality $\sum_{i}z_{i}en_{i}^{\text{b}}(\mathbf{r})=0$ (at least over spatial scales larger than the characteristic width of the Debye double layer and far away from the reservoir boundaries).  Then in the equation for the total current $\sum_{i}z_{i}e\mathbf{j}_{i}(\mathbf{r})=0$, the terms originating from gradients in the concentration cancel, leading to
\begin{equation}
 \left(\sum_{i=1}^{N}z_{i}en_{i}^{\text{b}}(\mathbf{r}) Q_{i}^{*}\right)\frac{\nabla T}{T}-\left(\sum_{i=1}^{N}z_{i}^{2}e^{2} n_{i}^{\text{b}}(\mathbf{r})\right)\mathbf{E}^{\text{th}}(\mathbf{r})=0.
\end{equation}
 To linear order in the thermal gradient, we replace the ion concentration and temperature by their unperturbed values $n_i^b(\mathbf{r}) \mapsto n_{i,0}^b, T(\mathbf{r}) \mapsto T_0$, such that the thermoelectric field becomes uniform
\begin{equation}\label{Eq:Electric_field_bulk}
 \mathbf{E}^{\text{th}}=-
\phi^{\text{th}} \frac{\nabla T}{T_0},   
\end{equation}
where we define the thermoelectric potential as
\begin{equation}
\phi^{\text{th}} =-\frac{\sum_{i=1}^{N}z_{i}n_{i,0}^{b}Q_{i}^{*}}{\sum_{i=1}^{N}z_{i}^{2}en_{i,0}^{b}}.
\end{equation}
Substituting the obtained thermoelectric field in Eq.~\eqref{Eq:bulk_current}, the steady state of the ionic solutes in bulk is governed to linear order in the thermal gradients  by
\begin{equation}\label{Eq:density_gradient_bulk}
 \frac{\nabla n_{i}^{\text{b}}(\mathbf{r})}{n_{i,0}^{\text{b}}}=-\frac{Q_{i}^{*} + q_{i}\phi^{\text{th}}}{k_B T_0} \frac{\nabla T}{T_0} =: -S_T^i \nabla T.
\end{equation}
From the last identity we read off  the  \emph{ionic Soret coefficients}
\begin{equation}\label{Eq:Solute_Soret}
 S_{T}^{i}= 
\mathcal{S}_{T}^{i}+ \frac{z_{i}e\phi^{\text{th}}}{k_{\text{B}} T_{0}^{2}},
\end{equation}
with $\mathcal{S}_{T}^{i}=Q_{i}^{*}/k_{\text{B}}T_0^2$. The first contribution arises  from hydration effects of the water molecules  and is connected to the ionic heat of transport $Q_{i}^{*}$, whereas the second contribution originates from electric migration in the thermoelectric field $\mathbf{E}^{\text{th}}$.

\section{Chebyshev spectral collocation}\label{App:Chebyshev_collocation}

Using the nonlinear transformation $\Phi_{L}(t)$, the derivatives with respect to the new variable $t$ are readily calculated as
\begin{subequations}
\begin{align}
\frac{\textrm{d}\phi_{0}(r)}{\textrm{d}r}&=\frac{(t-1)^2}{2L}\frac{\textrm{d}\phi_{0}(t)}{\textrm{d}t}, \\
\frac{\textrm{d}^{2}\phi_{0}(r)}{\textrm{d}r^{2}}&=\frac{(t-1)^4}{4L^2}\frac{\textrm{d}^{2}\phi_{0}(t)}{\textrm{d}t^{2}}+\frac{(t-1)^3}{2L^{2}}\frac{\textrm{d}\phi_{0}(t)}{\textrm{d}t},
\end{align}
\end{subequations}
by successively applying the chain rule. Consequently, the transformed nonlinear differential equation with respect to the variable $t\in\left[-1,1\right)$ reads
\begin{align}\label{Eq:transformed_PB_equation}
 \frac{(t-1)^4}{4L^2}&\frac{\textrm{d}^2\phi_{0}(t)}{\textrm{d}t^2}+\left[\frac{(t-1)^4(L-1)}{2L^{2}(L+1+t(L-1))}\right]\frac{\textrm{d}\phi_{0}(t)}{\textrm{d}t} \nonumber \\
 &+\kappa_{0}^{2}\sum_{i=1}^{N}z_{i}n_{i,0}^{\text{b}}\exp(-z_{i}\phi_{0}(t))=0.
\end{align}
and the boundary conditions transforms to
\begin{subequations}
\begin{align}
\lim\limits_{t\rightarrow 1}{\phi_{0}(t)}&=0, \\
 \frac{\textrm{d}\phi_{0}(t)}{\textrm{d}t}\bigg|_{t=-1}&=-\frac{L}{2}\sigma_{0}.
\end{align}
\end{subequations}
The solution $\phi_{0}(t)$ is approximated
at the Chebyshev-Gauss-Lobatto nodes
\begin{equation}
 t_{j}=\cos\left(\frac{j\pi}{M}\right),\quad j= 0,\ldots,M,
\end{equation}
by a global polynomial interpolant
\begin{equation}\label{Eq:pol_interpolant}
 \phi_{0}(t)\approx p_{M}(t)=\sum_{k=0}^{M}\varphi_{k}\ell_{k}(t),
\end{equation}
where $\varphi_{k}:=\phi_{0}(t_{k})$ and $\ell_{j}(t)$ denotes the Lagrange polynomial basis functions satisfying $\ell_{k}(t_{j})=\delta_{jk}$~\cite{Peyret:SMIVF:2002, Canuto:SMFD:1988}.
Then, the approximation of the $p$-th derivative of the function $\phi_{0}(t)$ is achieved by differentiating the interpolant and evaluating the result at the nodal points $\{t_{j}\}$ defining the Chebyshev differentiation matrices $\mathsf{D}^{(p)}$ with entries
\begin{equation}
\mathsf{D}_{jk}^{(p)}=\frac{\textrm{d}^{p}\ell_{k}(t)}{\textrm{d}t^{p}}\bigg|_{t=t_{j}}.
\end{equation}
For the first-order differentiation matrix $\mathsf{D}^{(1)}$, this yields~\cite{Boyd:CFSM:2003, Canuto:SMFD:1988} 
\begin{equation}\label{Eq:Chebyshev_matrix}
\mathsf{D}_{jk}^{(1)}=\begin{dcases}
							\frac{c_{j}}{c_{k}}\frac{(-1)^{j+k}}{t_{j}-t_{k}}, &  j\neq k \\
							-\sum_{k=0, k\neq j}^{M}\mathsf{D}_{jk}^{(1)},   &  j=k,
						 \end{dcases}
\end{equation}
where $ j,k=0,\ldots, M$, $c_{0}=c_{M}=2$ and $c_{l}=1$ for $l=1,\ldots, M-1$. Here, we reduce possible cancellation errors in the diagonal elements of the differentiation matrix as $M$ increases by calculating them from the analytic expressions for the off-diagonal elements~\cite{Baltensperger:CMA_37:1999, Bayliss:JCP_116:1995, Baltensperger:SIAMJSC_24:2003} (see first line in Eq.~\eqref{Eq:Chebyshev_matrix}). Furthermore, the summands in Eq.~\eqref{Eq:Chebyshev_matrix} are rearranged in  ascending order to avoid smearing. Moreover, the second-order Chebyshev differentiation matrix can be obtained from $\mathsf{D}^{(2)}=\left(\mathsf{D}^{(1)}\right)^{2}$, applying the same correction technique for the diagonal entries $\mathsf{D}_{jj}^{(2)}$ which leads to significantly higher accuracy. Consequently, the numerical differentiation at the Chebyshev collocation points $t_{j}$ can be written in vector form 
\begin{equation}
 \mathbf{y}^{(p)}= \mathsf{D}^{(p)}\mathbf{y},\quad p=1,2,
\end{equation}
with the coefficient vectors
\begin{subequations}
\begin{align}
 \mathbf{y}&:=\left(\varphi_0,\ldots,\varphi_{M}\right)^{T}, \\
 \mathbf{y}^{(p)}&:=\left(\varphi^{(p)}_{0},\ldots,\varphi^{(p)}_{M}\right)^{T}.
\end{align}
\end{subequations}
The collocation method states that the polynomial interpolant [Eq.~\eqref{Eq:pol_interpolant}] satisfies the nonlinear ODE [Eq.~\eqref{Eq:transformed_PB_equation}] at the inner collocation points ${t_{j}}, \; j=1,\ldots,M-1$, yielding the discrete approximation
\begin{align}
\sum_{k=0}^{M}&\left[\frac{(t_{j}-1)^4(L-1)}{2L^2(L+1+t_{j}(L-1))}\textsf{D}^{(1)}_{jk}+\frac{(t_{j}-1)^4}{4L^2}\textsf{D}^{(2)}_{jk}\right]\varphi_{k} \nonumber\\
&\quad+\kappa_0^{2}\sum_{i=1}^{N}z_{i}n_{i,0}^{\text{b}}\exp\left(-z_{i}\varphi_{k}\right)=0.
\end{align}
Evaluating the boundary conditions at the noncollocated endpoints $t_{0}$ and $t_{M}$
\begin{subequations}
\begin{align}
\varphi_{0}&=0, \\
\sum_{k=0}^{M}\textsf{D}_{Mk}^{(1)}\varphi_{k} &= -\frac{L}{2}\sigma_{0}
\end{align}
\end{subequations}
results in a set of $M+1$ nonlinear algebraic equations for the variables $\varphi_{1}, \ldots, \varphi_{M}$ which are solved using a Newton-Raphson method with a constant initial guess $\varphi_{j}=1$ for all $j= 1,\ldots, M$.
An approximate solution on the unbounded domain $[1,\infty)$ in terms of a transformed barycentric interpolant then reads
\begin{equation}
\phi_{0}(r)\approx p_{M}(r)=\dfrac{\sum_{j=0}^{M}W_{j}(r)\varphi_{j}}{\sum_{j=0}^{M}W_{j}(r)},
\end{equation}
with $W_{j}(r)=(-1)^{j}w_{j}/\left[\Phi_{L}^{-1}(r)-\Phi_{L}^{-1}(r_{j})\right]$ and the reduced barycentric weights $w_{0}=w_{M}=2$ or $w_{j}=1,\, j=1,\ldots, M-1$~\cite{Berrut:SIAMR_46:2004}. Similar expressions for the first and second derivative can be obtained by substituting $\phi_{j}$ with $\textsf{D}^{(1)}\phi_{j}$, respectively $\textsf{D}^{(2)}\phi_{j}$.

\section{Matrix representation for the asymptotic constants}\label{App:Matrix_rep_asymp_sol}

\renewcommand{\arraystretch}{1.2}
\begin{table*}[t]
\centering
\begin{tabular}{lSSSSSS}
\toprule
\textbf{Ion} &  $\text{K}^{+}$ & $\text{Na}^{+}$ & $\text{TRIS-H}^{+}$ & $\text{Cl}^{-}$ & $\text{F}^{-}$ & $\text{OH}^{-}$ \\
\midrule
$\mathcal{S}_{T}^{i}\,\left[10^{-4}\,\text{K}^{-1}\right]$ & 35.1 & 46.9 & N/A & 7.18 & 53.2 & 233 \\
$\mu_{i}^{e}\,\left[10^{-8}\,\text{m}^{2}\text{V}^{-1}\text{s}^{-1}\right]$ & 7.62 & 5.19 & N/A & -7.91 & -5.74 & -20.5 \\
$\text{Pe}_{i}$ & 0.263 & 0.385 & N/A & 0.253 & 0.349 & 0.0976 \\
\midrule
\textbf{salt} & KCl & NaOH & $\text{TRIS-HCl}$ & KF & NaF &  \\
\midrule
$\phi^{\text{th}}$ & -0.416 & 2.77 & N/A & 0.270 & 0.0939 &  \\
   \bottomrule
\end{tabular}
\caption{Typical values for the different parameters used to determine the numerical predictions in Figs.~\ref{fig:plot_1}-\ref{fig:plot_5} for different monovalent ions and the corresponding salts immersed in an infinitely dilute aqueous solution. Ionic heats of transport to calculate $\mathcal{S}_{T}^{i}=Q_{i}^{*}/k_{\text{B}}T_{0}^2$ have been taken from~\cite{Takeyama:JSC_17:1988} and the electric mobilities are converted from ion limiting condictivities~\cite{Rumble:CRC_55:2021} at temperature $T_{0}=298.15\,\text{K}$ ($25\,^{\circ}\text{C}$). P{\'e}clet numbers $\text{Pe}_{i}$ and thermoelectric potentials $\phi^{\text{th}}$ are calculated from these values. Concerning the TRIS-HCl buffer, relevant numbers are mentioned in the text.}
\label{tab:table_1}
\end{table*}
\renewcommand{\arraystretch}{1}

From the solutions $\omega_i^k(r),R^k(r),\, k=1,\ldots, N+2$ for the $N+2$ linear ODEs we can calculate the components of the coefficient matrix $\mathsf{A}$ for the linear problem [Eq.~\eqref{Eq:linear_problem_constants}] 
\begin{subequations}
\begin{align}
\mathsf{A}_{i,k}&=\frac{\textrm{d}\omega_{i}^{k}(r)}{\textrm{d}r}\bigg|_{r=1}, \\
\mathsf{A}_{N+1,k}&=R^{k}(r)|_{r=1}, \\
\mathsf{A}_{N+2,k}&=\frac{\textrm{d}R^{k}(r)}{\textrm{d}r}\bigg|_{r=1}-\lambda \frac{\textrm{d}^{2}R^{k}(r)}{\textrm{d}r^{2}}\bigg|_{r=1},
\end{align}
\end{subequations}
with $i=1,\ldots, N$ and $k=1,\ldots, N+2$. 
In addition, the components of the corresponding vector $\mathbf{B}$  are given as
\begin{subequations}
\begin{align}
B_{i}=&-\frac{\textrm{d}\omega_{i}(r)}{\textrm{d}r}\bigg|_{r=1} \nonumber \\
&- \begin{cases}
0, & (1) \\
z_{i}\left[\phi_{0}(r) + \phi^{\text{th}}\right]|_{r=1}, & (2)
\end{cases}, \\
B_{N+1}=&- \frac{R(r)}{r}|_{r=1} + \begin{cases}
1/2, & (1) \\
0, & (2)
\end{cases}, \\
B_{N+2}=&
-\frac{\textrm{d}R(r)}{\textrm{d}r}\bigg|_{r=1}+\lambda\frac{\textrm{d}^{2}R(r)}{\textrm{d}r^{2}}\bigg|_{r=1} \nonumber \\
& + 
\begin{cases}
1/2, & (1) \\
0, & (2)
\end{cases}
\end{align}
\end{subequations}
with $i=1,\ldots, N$. Thus the asymptotic coefficients for each problem can be calculated formally as
\begin{equation}
\mathbf{C}=\mathsf{A}^{-1} \cdot \mathbf{B}.
\end{equation}

\section{Typical values for relevant parameters}
\label{App:Relevant_parameter_values}

In this appendix, we provide typical values of the various parameters for an aqueous electrolyte in the presence of different salt ions. Unless otherwise stated all values are determined at reference temperature $T_{0}=298.15\,\text{K}$ ($25\,^{\circ}\text{C}$).
Here the solvent is modeled as pure water with relative dielectric permittivity $\epsilon_{\text{r}}^{0}=78.304$~\cite{Malmberg:JRNBS_1:1956}, logarithmic derivative $\alpha=1.35$~\cite{Rumble:CRC_55:2021} and solvent viscosity $\eta=890.45\times 10^{-6}\,\text{Pa s}$. In addition, Soret coefficients $\mathcal{S}_{T}^{i}$ arising from hydration effects of the water molecules, electric mobilities $\mu_{i}^{e}=z_{i}e\mu_{i}^{0}$ and the corresponding ionic P{\'e}clet numbers $\text{Pe}_{i}$ for different ion species  are summarized in Table~\ref{tab:table_1} and refer to an infinitely dilute aqueous solution. We also list the dimensionless thermoelectric potential $\phi^{\text{th}}$ for the various monovalent salts. It can be calculated as $\phi^{\text{th}}=-(\mathcal{S}_{T}^{+}-\mathcal{S}_{T}^{-})T_{0}/2$ from $\mathcal{S}_{T}^{\pm}=Q_{\pm}^{*}/k_{\text{B}}T_{0}^2$  for cations ($+$) and anions ($-$) arising from the heat of ion hydration $Q_{\pm}^{*}$, which had been measured experimentally by~\cite{Takeyama:JSC_17:1988} at temperature $T_{0}$ for a broad range of different ionic solutes. Again, since relevant values for the $\text{TRIS-H}^{+}$ are not available, we follow Ref.~\cite{Burelbach:EPJE_42:2019} and choose $\mathcal{S}_{T}^{\text{TRIS-H}^{+}}=\mathcal{S}_{T}^{\text{Na}^{+}}$ together with the mobility $\mu_{\text{TRIS-H}^{+}}^{e}=2.67\times 10^{-8}\,\text{m}^{2}\text{V}^{-1}\text{s}^{-1}$ taken from a similar organic compound, the amino acid Leucine~\cite{Wronski:JCA_657:1993}. All other mobilities are converted from limiting equivalent conductivities of the ions~\cite{Rumble:CRC_55:2021}. Moreover, using the Stokes-Einstein relation, the ionic P{\'e}clet numbers are computed from $\text{Pe}_{i}=U_{0}a/D_{i}^{0}=\epsilon_{0}\epsilon_{\text{r}}^{0}k_{\text{B}}T_{0}z_{i}/\eta\mu_{i}^{e}$ where only properties of the dissolved ions except for the solvent viscosity determine their values. 


%\bibliographystyle{apsrev}
%\bibliographystyle{apsrev4-1}
\bibliography{Summary_thermophoresis_final.bib}

\end{document}